\newcommand \Mpc {h^{-1}{\rm Mpc}}
\newcommand \kpc {h^{-1}{\rm kpc}}

\newcommand \farcm{\hbox{$.\!\!^{\prime}$}}
\newcommand \arcm{\hbox{$^{\prime}$}}
\newcommand \arcs{\hbox{$^{\prime\prime}$}}

\newcommand \kms {{\rm km~s}^{-1}}
\newcommand \msun {h^{-1} M_\odot}
\newcommand \beqn {\begin{equation}}
\newcommand \eeqn {\end{equation}}
\newcommand \ncztot {15665 } 
\newcommand \nczcairns {3471 } 
\newcommand \nczpub {1621 } 
\documentclass[12pt,preprint]{aastex}
\usepackage{epsfig}
\usepackage{natbib}

\begin{document}

\title{CAIRNS: The Cluster And Infall Region Nearby Survey
I. Redshifts and Mass Profiles}

\author{Kenneth Rines\altaffilmark{1}, Margaret J. Geller\altaffilmark{2},
Michael J. Kurtz\altaffilmark{2}, and Antonaldo Diaferio\altaffilmark{3}} 
\email{krines@cfa.harvard.edu}

\altaffiltext{1}{Harvard-Smithsonian Center for Astrophysics, 60 Garden St,
MS 10, Cambridge, MA 02138 ; krines@cfa.harvard.edu}
\altaffiltext{2}{Smithsonian Astrophysical Observatory; mgeller, mkurtz@cfa.harvard.edu}
\altaffiltext{3}{Universit\`a degli Studi di Torino,
Dipartimento di Fisica Generale ``Amedeo Avogadro'', Torino, Italy; diaferio@ph.unito.it}

\begin{abstract}

The CAIRNS (Cluster And Infall Region Nearby Survey) project is a
spectroscopic survey of the infall regions surrounding eight nearby,
rich, X-ray luminous clusters of galaxies.  We collect \ncztot
redshifts (\nczcairns new or remeasured) within $\sim 5-10\Mpc$ of the
centers of the clusters, making it the largest study of the infall
regions of clusters.  We determine cluster membership and the mass
profiles of the clusters based on the phase space distribution of the
galaxies.  All of the clusters display decreasing velocity
dispersion profiles.  The mass profiles are fit well by functional
forms based on numerical simulations but exclude an isothermal sphere.
Specifically, NFW and Hernquist models provide good descriptions of
cluster mass profiles to their turnaround radii.  Our sample shows
that the predicted infall pattern is ubiquitous in rich, X-ray
luminous clusters over a large mass range. The caustic mass estimates
are in excellent agreement with independent X-ray estimates at small
radii and with virial estimates at intermediate radii.  The mean ratio
of the caustic mass to the X-ray mass is $1.03\pm0.11$ and the mean
ratio of the caustic mass to the virial mass (when corrected for the
surface pressure term) is $0.93\pm0.07$.  We further demonstrate that
the caustic technique provides reasonable mass estimates even in
merging clusters.

\end{abstract}


\section{Introduction}

Clusters of galaxies are the most massive gravitationally relaxed
systems in the universe.  They offer a unique probe of the properties
of galaxies and the distribution of matter on intermediate scales.
The dynamically relaxed centers of clusters are surrounded by infall
regions in which galaxies are bound to the cluster but are not in
equilibrium.  These galaxies populate a regime between that of relaxed
cluster cores and the surrounding large-scale structure where the
transition from linear to non-linear clustering occurs.  Recently,
various investigators have explored infall regions using two-body
dynamics of binary clusters \citep[e.g.][]{mw97}, the virial theorem
in superclusters \citep{small98}, weak lensing \citep{kaiserxx}, and
caustics in redshift space
\citep{gdk99,rqcm,rines2000,kg2000,drink,tustin,rines01a,rines01b, rines02,bg03}.  

In redshift space, the infall regions of clusters form a
characteristic trumpet-shaped pattern.  The presence of sharp features
in redshift space near clusters in early redshift surveys hinted at
the existence of such a pattern \citep{kg82,dgh86,ostriker88}.  These
features arise because galaxies fall into the cluster as the cluster
potential overwhelms the Hubble flow \citep{kais87,rg89}. Under simple
spherical infall, the galaxy phase space density becomes infinite at
the location of these features, which were therefore termed
caustics. \citet[][hereafter DG]{dg97} and \citet[][hereafter
D99]{diaferio1999} analyzed the dynamics of infall regions with
numerical simulations and found that in the outskirts of clusters,
random motions due to substructure and non-radial motions make a
substantial contribution to the amplitude of the caustics which
delineate the infall regions. DG showed that the amplitude of the
caustics is a measure of the escape velocity from the cluster;
identification of the caustics therefore allows a determination of the
mass profile of the cluster on scales $\lesssim 10\Mpc$.

DG and D99 show that nonparametric measurements of caustics yield
cluster mass profiles accurate to $\sim$50\% on scales of up to 10
$h^{-1}$ Mpc.  This method assumes only that galaxies trace the
velocity field. Indeed, simulations suggest that little or no velocity
bias exists on linear and mildly non-linear scales
\citep{kauffmann1999a,kauffmann1999b}. \citet{vh98} used  
simulations to explore an alternative parametric analysis of the
infall region using a maximum likelihood method. Their technique
requires assumptions about the functional forms of the density profile
and the velocity dispersion profile.  

We discuss the Cluster And Infall Region Nearby Survey (CAIRNS),
a redshift survey of the infall regions of 8 nearby galaxy
clusters.   Previous papers from this survey examine A576
\citep{rines2000}, A2199/A2197 \citep{rines01b,rines02}, and A1656
\citep[Coma;][]{gdk99,rines01a}.   We include new or remeasured
redshifts for \nczcairns galaxies in the infall regions of six of the
eight clusters in this survey. 

The CAIRNS project tests whether the caustic pattern described in DG
and D99 is common in nearby rich clusters and thus evaluates the
feasibility of measuring cluster mass profiles at large radii from
redshift surveys using the caustic technique.  Other goals of CAIRNS
include (1) measuring the mass-to-light ratio as a function of scale
\citep{rines2000,rines01a}, (2) detecting substructures in infall
regions as a probe of structure formation \citep{rines01b,rines02},
and (3) studying the dependence of the spectroscopic properties of
galaxies on environment over a large range of densities.

CAIRNS also provides an important zero-redshift benchmark for
comparison with more distant systems \citep[e.g.,][]{ellingson01}.
The CNOC1 project assembled an ensemble cluster from X-ray selected
clusters at moderate redshifts. The CNOC1 ensemble cluster samples
galaxies up to $\sim$2 virial radii \citep[see][and references
therein]{cye97,ellingson01}.  The caustic pattern is easily visible in
the ensemble cluster, but \citet{cye97} apply only Jeans analysis to
the cluster to determine an average mass profile.  Recently,
\citet{bg03} analyzed cluster redshifts from the 2dF 100,000 redshift
data release. They stacked 43 poor clusters to produce an ensemble
cluster containing 1345 galaxies within 2 virial radii and analyzed
the properties of the ensemble cluster with both Jeans analysis and
the caustic technique.  \citet{bg03} find good agreement between the
two techniques; the caustic mass profile beyond the virial radius
agrees well with an extrapolation of the Jeans mass profile.  In
contrast to these studies, the CAIRNS clusters are sufficiently well
sampled to apply the caustic technique to the individual clusters.

We describe the cluster sample in $\S$ 2.  We describe the
spectroscopic observations and the redshift catalogs in $\S$ 3.  In
$\S$ 4, we review the caustic technique and use it to estimate the
cluster mass profiles. We compare the caustic mass profiles to simple
parametric models in $\S$ 5.  We compute the velocity dispersion
profiles in $\S$ 6.  We compare the caustic mass profiles to X-ray and
virial mass estimators in $\S$ 7.  We discuss our results and conclude
in $\S 8$.  We assume $H_0 = 100 h~\kms, \Omega _m = 0.3, 
\Omega _\Lambda = 0.7$ throughout.

\section{The CAIRNS Cluster Sample}

We selected the CAIRNS parent sample from all nearby
($cz_\odot<15,000~\kms$), Abell richness class $R\geq1$
\citep[][]{aco1989}, X-ray luminous ($L_X>2.5 \times 10^{43}
h^{-2}$erg s$^{-1}$) galaxy clusters with declination
$\delta>-15^\circ$. Using X-ray data from the X-ray Brightest Abell
Clusters (XBACs) catalog \citep{xbacs, xbacserr}, the parent cluster
sample contains 14 systems.  We selected a representative sample of 8
of these 14 clusters (Table
\ref{sample}).  The redshifts and X-ray properties listed in Table
\ref{sample} are from \citep{xbacs}, the velocity dispersions are from
$\S 3.2$, and the richness classes are from \citet{aco1989}.  The 6
clusters meeting the selection criteria but not targeted in CAIRNS
because of limited observing time are: A193, A426, A2063, A2107,
A2147, and A2657.  The 8 CAIRNS clusters span a variety of
morphologies, from isolated clusters (A496, A2199) to major mergers
(A168, A1367).  We also do not include AWM7, which meets all the
requirements for inclusion but is not in the Abell catalog.
\citet{kg2000} describe a large spectroscopic survey of AWM7.

The redshift and declination limits are set by the small aperture and
the location of the 1.5-m Tillinghast telescope used for the vast
majority of our spectroscopic observations. The richness minimum
guarantees that the systems contain sufficiently large numbers of
galaxies to sample the velocity distribution.  The X-ray luminosity
minimum guarantees that the systems are real clusters and not
superpositions of galaxy groups
\citep[cf. the discussion of A2197 in][]{rines01b, rines02}.  
Three additional clusters with smaller X-ray luminosities (A147, A194
and A2197) serendipitously lie in the survey regions of A168 and
A2199.  A147 and A2197 lie at nearly identical redshifts to A168 and
A2199; their dynamics are probably dominated by the more massive
cluster \citep{rines02}.  A194, however, is cleanly separated from
A168 and we therefore analyze it as a ninth system.  The inclusion of
A194 extends the parameter space covered by the CAIRNS sample.  The
X-ray temperature of A194 listed in \citep{xbacs} is an extrapolation
of the $L_X - T_X$ relation; in Table \ref{sample} we therefore list the
direct temperature estimate of \citet{1998PASJ...50..187F} from {\em
ASCA} data. \citet{1998PASJ...50..187F} lists X-ray temperatures for 6
of the 8 CAIRNS clusters which agree with those listed in
\citet{xbacs}. 

\section{Observations}

\subsection{Spectroscopy}

We have collected \ncztot redshifts within a radius of $\sim$10$\Mpc$
of the 8 clusters in the CAIRNS sample.  Of this total, \nczcairns are
new or remeasured and \nczpub were published in \citet{rines2000} and
\citet{rines02}.  New and remeasured redshifts were obtained with the
FAST spectrograph \citep{fast} on the 1.5-m Tillinghast telescope of
the Fred Lawrence Whipple Observatory (FLWO).  FAST is a high
throughput, long slit spectrograph with a thinned, backside
illuminated, antireflection coated CCD detector.  The slit length is
180\arcs; our observations used a slit width of 3\arcs ~and a 300
lines mm$^{-1}$ grating.  This setup yields spectral resolution of 6-8
\AA ~and covers the wavelength range 3600-7200 \AA.  We obtain
redshifts by cross-correlation with spectral templates of
emission-dominated and absorption-dominated galaxy spectra created
from FAST observations \citep{km98}.  The typical uncertainty in the
redshifts is 30$~\kms$.

To construct the primary target catalogs for all clusters except
A1367, we selected targets from digitized images of the POSS I 103aE
(red) plates.  These catalogs are roughly complete to $E=15-16$,
although the star-galaxy classification and photometric uncertainties
limit the completeness of these catalogs. We selected A1367 targets
from the Zwicky catalog and the Automated Plate
Scanner$\footnote{http://aps.umn.edu}$ catalog of the POSS I O (blue)
plates. The A1367 catalog is complete to $m_{Zw}$=15.7 within a radius
of 6$^\circ$ of the center of A1367. \citet{rines02} discuss
incompleteness in the APS catalog.

 Where possible, we use CCD photometry to define more complete and
accurate catalogs.  Our redshift survey of A576 \citep{rines2000} is
complete to Kron-Cousins R=16.5 in a $3^\circ \times 3^\circ$ square,
and our redshift survey of Coma \citep{rines01a} is complete to
K$_s$=12.2 within a radius of $8^\circ$ based on 2MASS isophotal
magnitudes (K$_s$=20 mag arcsec$^{-2}$ isophote).

An important difference between the FAST spectra collected for CAIRNS
and those collected for other, larger redshift surveys
\citep{2df,sdss} is that CAIRNS suffers no incompleteness due to fiber
placement constraints.
Another difference is that the long-slit FAST spectra sample light
from larger fractions of galaxy areas than fiber spectra.  Thus, the
effects of aperture bias \citep[e.g.,][]{apbias} on spectral
classification are greatly reduced.  \citet{bcarter} show that a
spectroscopic survey of field galaxies obtained with identical
instrumentation in a similar redshift range contains no significant
aperture bias.

\subsection{Redshift Catalogs}

We compiled redshifts from the literature as collected by the
NASA/IPAC Extragalactic Database (NED\footnote{The NASA/IPAC
Extragalactic Database is available at
http://nedwww.ipac.caltech.edu/index.html}).  We then matched the
catalogs with a $10\arcs$ search radius.  When an object has a
redshift from both FAST and NED, we select the FAST redshift for the
sake of homogeneity.  We inspect all matches with redshift differences
of $|\Delta v| \ge 100~\kms$ to determine whether the redshifts are
for the same object (close pairs of galaxies can be erroneously
matched to each other). When the targets are identical, we adopt the
FAST redshift unless the cross-correlation R value \citep{km98} is
$\le 3.0$, in which case we adopt the NED redshift.
The sources of the literature redshifts are too numerous to list here.
A119, A168, and A194 all lie in or near the first SDSS survey strip;
the SDSS Early Data Release contributes a plurality of the literature
redshifts.

We apply the prescription of \citet{danese} to determine the mean
redshift $cz_\odot$ and projected velocity dispersion $\sigma_p$ of
each cluster from all galaxies within 1.5~$\Mpc$ (1 Abell radius or
$R_A$) of its X-ray center listed in Table \ref{sample}.  We calculate
$\sigma_p$ in two different ways.  3$\sigma$ clipping, an iterative
process where we first calculate the mean and variance of the
distribution, then remove all galaxies offset by more than 3$\sigma$
from the mean, and repeat as necessary.  The second technique is use
of the caustics to define membership (see $\S$ 4.2).  We then
calculate $\sigma_p$ using only the cluster members projected within
$r_{200}$, the radius which encloses a mass density 200 times the
critical density.  Note that our estimates of $r_{200}$ do not depend
on $\sigma_p$.  We list both estimates of $\sigma_p$ in Table
\ref{sample}; the values of $\sigma_p$ calculated from the caustic
members are all smaller than those calculated with 3$\sigma$ clipping.
This result suggests that the caustic membership criterion is more
restrictive and less sensitive to interlopers than 3$\sigma$ clipping
(see $\S$ 4.2).  For consistency, we adopt the projected velocity
dispersions from the caustic technique in all further analyses.

Table \ref{catalogs} describes the catalog areas and lists the number
of redshifts $N_{cz}$ in each catalog, the number $N_{CAIRNS}$
obtained with FAST, and the number $N_{mem}$ of cluster members inside
the caustics and projected within $r_{200}$.
Note that A119, A168, and A194 overlap significantly; the number of
unique redshifts in the combined catalog of A119 and A168 (Table
\ref{a119a168cz}) is 5055 (855 new or remeasured).  For each galaxy,
Table \ref{a119a168cz} indicates which cluster is less than 5$^\circ$
away or ``Both'' if both A119 and A168 are closer than 5$^\circ$.
Note that this designation is not a membership classification.  Tables
\ref{a194cz}, \ref{a496cz}, \ref{a539cz}, \ref{a1367cz}, and \ref{comacz}
contain the redshift catalogs for A194, A496, A539, A1367, and Coma
respectively.  Note that the redshift catalog for A194 includes only
literature redshifts (2431 redshifts from the literature are included
in Table \ref{a194cz} but are not in Table \ref{a119a168cz}) along
with some FAST redshifts also listed in Table \ref{a119a168cz}.  The
redshift catalog of Coma supersedes the online catalog referred to in
\citet{rines01a}.  Redshift catalogs for A576 and A2199 are available
in \citet{rines2000} and \citet{rines02}.  Because the redshift
catalogs include substantial contributions from the literature, the
sampling and completeness are not uniform.  Table \ref{catalogs}
therefore cannot be used to compare cluster richnesses.
\citet{gdk99} show that imposing a magnitude limit on the redshift
catalog around Coma does not affect the location of the caustics or
the resulting mass profile.  Similarly, imposing a magnitude limit to
these redshift catalogs does not significantly affect any of our
results.

\subsection{The $\sigma_p -T_X$ Relation of CAIRNS Clusters}

\begin{figure}
\figurenum{1}
\plotone{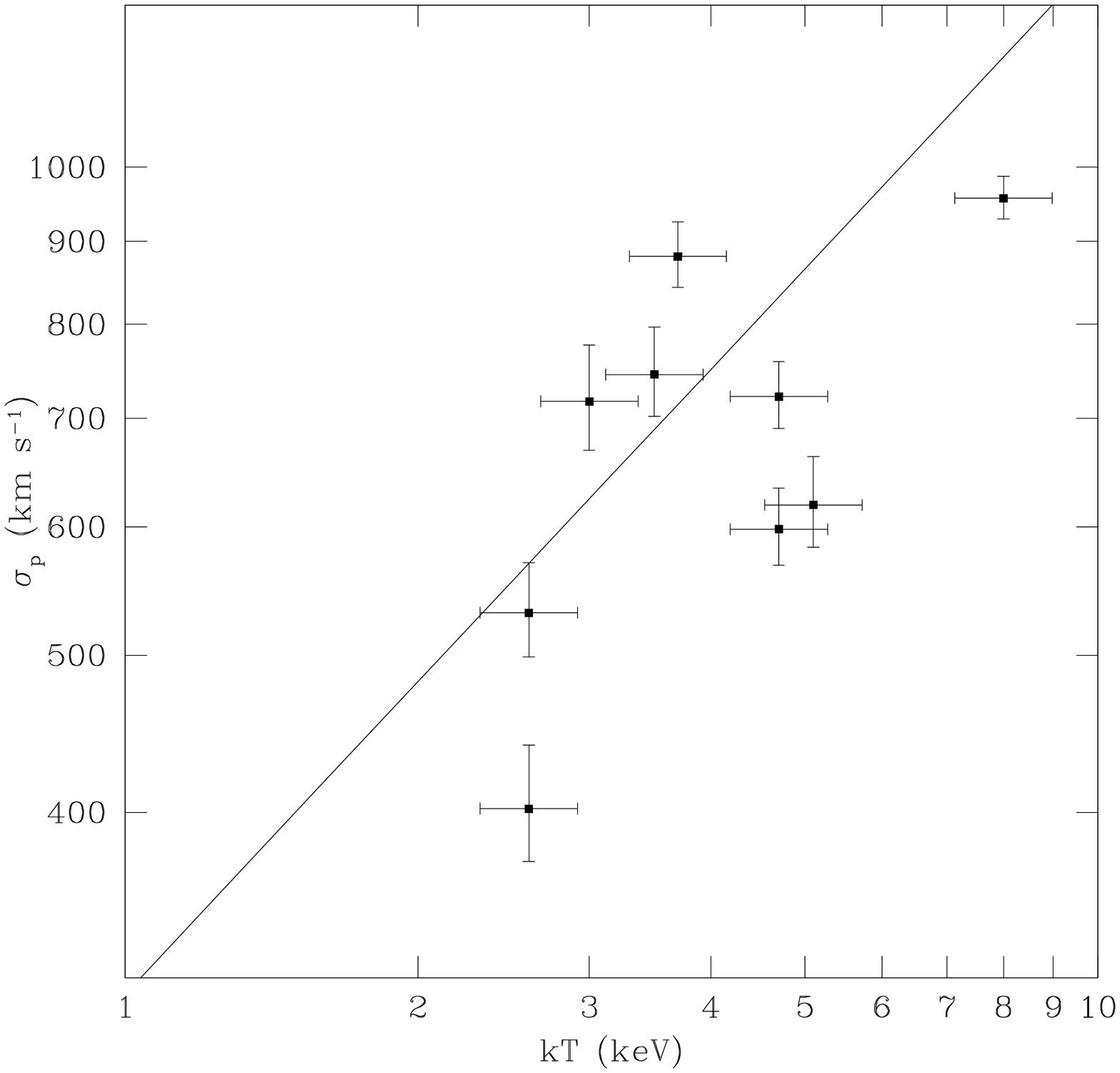}
\caption{\label{sigmat} Velocity dispersion versus X-ray temperature
for the CAIRNS clusters. The solid line shows the $\sigma -T_X$
relation of Xue \& Wu 2000.}  
\end{figure}

Figure \ref{sigmat} shows the $\sigma_p -T_X$ relation for the CAIRNS
clusters along with the best-fit $\sigma_p -T_X$ relation (using
orthogonal distance regression) of a large sample of groups and
clusters \citep{2000ApJ...538...65X}.  Although the scatter is large,
the CAIRNS clusters follow the same relation.

\section{Calculating the Mass Profiles}

\subsection{Method}

We briefly review the method DG and D99 developed to estimate the mass
profile of a galaxy cluster by identifying its caustics in redshift
space.  The method assumes that clusters form in a hierarchical
process.  Application of the method requires only galaxy redshifts
and sky coordinates.  Toy models of simple spherical infall onto
clusters produce sharp enhancements in phase space density.  These
enhancements, known as caustics, appear as a trumpet shape in scatter
plots of redshift versus projected clustercentric radius
\citep{kais87}.  DG and D99 show that random motions smooth out the
sharp pattern expected from simple spherical infall into a dense
envelope in the redshift-radius diagram \citep[see also][]{vh98}.  The
edges of this envelope can be interpreted as the escape velocity as a
function of radius.  Galaxies outside the caustics are also outside
the turnaround radius.  The caustic technique provides a well-defined
boundary between the infall region and interlopers; one may think of
the technique as a method for defining membership that gives the
cluster mass profile as a byproduct.

The amplitude $\mathcal{A} \mathnormal{(r)}$ of the caustics is half of
the distance between the upper and lower caustics in redshift
space. Assuming spherical symmetry, $\mathcal{A} \mathnormal(r)$ is
related to the cluster gravitational potential $\phi (r)$ by
\beqn
\mathcal{A} \mathnormal {^2 (r) = -2 \phi (r)\frac{1-\beta (r)}{3-2\beta (r)}}
\eeqn
where $\beta (r) = \sigma_t(r)/\sigma_r(r)$ is the velocity anisotropy
parameter and $\sigma_t$ and $\sigma_r$ are the tangential and radial
velocity dispersions respectively. DG show that the mass of a
spherical shell of radii [$r_0,r$] within the infall region is the
integral of the square of the amplitude $\mathcal{A} \mathnormal{(r)}$
\beqn
GM(<r)-GM(<r_0) = F_\beta \int_{r_0}^r \mathcal{A} \mathnormal{^2(x)dx}
\eeqn
where $F_\beta \approx 0.5$ is a filling factor with a numerical value
estimated from simulations. Variations in $F_\beta$ lead to some
systematic uncertainty in the derived mass profile (see D99 for a more
detailed discussion).

Operationally, we identify the caustics as curves which delineate a
significant decrease in the phase space density of galaxies in the
projected radius-redshift diagram.  For a spherically symmetric
system, taking an azimuthal average amplifies the signal of the
caustics in redshift space and smooths over small-scale substructures.
We perform a hierarchical structure analysis to locate the centroid of
the largest system in each cluster.  We adaptively smooth the
azimuthally averaged data and choose a threshold phase space density.
The upper and lower caustics at a given radius are the redshifts at
which this threshold density is exceeded when approaching the central
redshift from the ``top'' and ``bottom'' respectively of the
redshift-radius diagram.  Because the caustics of a spherical system
are symmetric, we adopt the smaller of the upper and lower caustics as
the caustic amplitude $\mathcal{A} \mathnormal{(r)}$ at that radius.
This procedure reduces the systematic uncertainties introduced by
interlopers, which generally lead to an overestimate of the caustic
amplitude.  \citet{rines02} explore the effects of altering some of
these assumptions and find that the differences are generally smaller
than the estimated uncertainties. 

D99 described this method in detail and showed that, when applied to
simulated clusters containing galaxies modelled with semi-analytic
techniques, it recovers the actual mass profiles to radii of
5-10$~\Mpc$ from the cluster centers.  D99 give a prescription for
estimating the uncertainties in the caustic mass profiles.  The
uncertainties estimated using this prescription reproduce the actual
differences between the caustic mass profiles and the true mass
profiles.  The uncertainties in the caustic mass profiles of observed
clusters may be smaller than the factor of 2 uncertainties in the
simulations.  This difference is due in part to the large number of
redshifts in the CAIRNS redshift catalogs relative to the simulated
catalogs.  Furthermore, the caustics are generally more cleanly
defined in the data than in the simulations.  Clearly, more
simulations which better reproduce the appearance of observed caustics
and/or include fainter galaxies would be useful in determining the
limits of the systematic uncertainties in the caustic technique.  In
$\S 7$, we use the properties of the X-ray gas to determine the
scatter in the caustic mass estimates relative to X-ray mass
estimates.

\subsection{Caustics in the CAIRNS Clusters}

\begin{figure}
\figurenum{2}
\epsscale{0.8}
\plotone{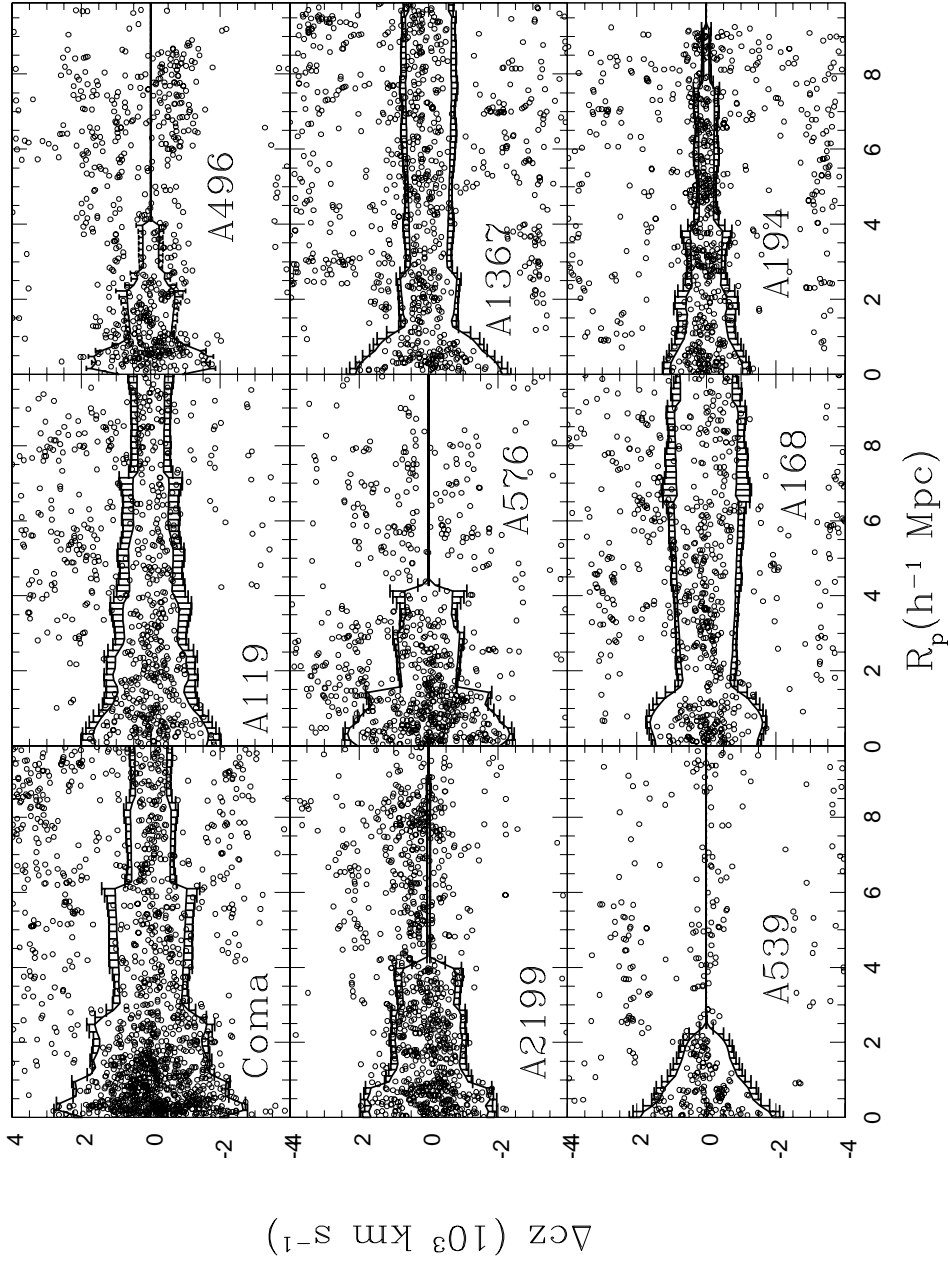}
\caption{\label{allcaustics} Redshift versus radius for galaxies around
the CAIRNS clusters. The caustic pattern is evident as the
trumpet-shaped regions with high density.  The solid lines indicate
our estimate of the location of the caustics in each cluster.  The
errorbars are 1-$\sigma$ uncertainties and are shown only on one side
of each caustic for clarity.}
\end{figure}

\begin{figure}
\figurenum{3}
\plotone{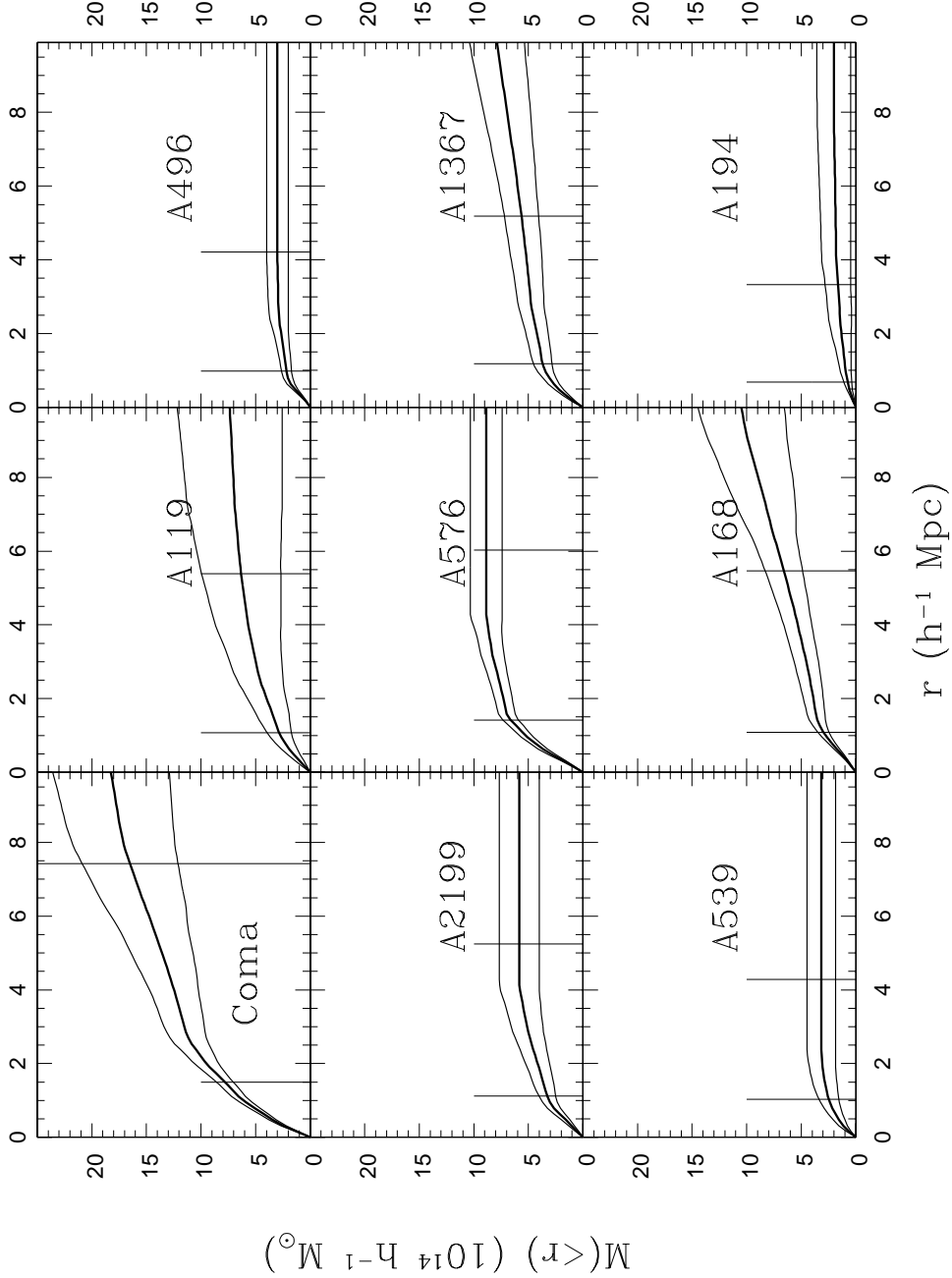} 
\caption{\label{allmp} Caustic mass profiles for the CAIRNS
clusters. The thick solid lines show the caustic mass profiles and the
thin lines show the 1-$\sigma$ uncertainties in the mass profiles. The
axes are identical in all panels. The vertical bars indicate $r_{200}$
and the turnaround radius.}
\end{figure}

Figure \ref{allcaustics} displays the projected radii and redshifts of
galaxies surrounding the CAIRNS clusters, ordered left to right and
top to bottom by decreasing X-ray temperature.  The expected caustic
pattern is easily visible in all systems; we calculate the shapes with
the technique described in D99 using a smoothing parameter of $q=25$.
Previous investigations show that the mass profiles are insensitive to
changes in the smoothing parameter \citep{gdk99,rines2000,rines02}.
Table \ref{centers} lists the hierarchical centers.  These centers
generally agree with the X-ray positions (Table \ref{sample}) with a
median difference of 56$~\kpc$ and with the redshift centers as
determined using galaxies within $R_A$ (Table \ref{catalogs}) with a
median difference of -0.0002.  Figure \ref{allcaustics} shows the
caustics and Figure \ref{allmp} shows the associated mass profiles.
Note that the caustics extend to different radii for different
clusters.  D99 show that the appearance of the caustics depends
strongly on the line of sight; projection effects can therefore
account for most of the differences in profile shape seen in Figure
\ref{allcaustics} without invoking non-homology among clusters.
We use the caustics to determine cluster membership.  Here, the term
``cluster member'' refers to galaxies both in the virial region and in
the infall region.  Figure \ref{allcaustics} shows that the caustics
effectively separate cluster members from background and foreground
galaxies, although some interlopers may lie within the caustics. This
clean separation affirms our adoption of velocity dispersions
calculated from cluster members as defined by the caustics ($\S$3.2).

In the simulations of D99, the degree of definition of the caustics
depends on the underlying cosmology; caustics are better defined in a
low-density universe than a closed, matter-dominated universe.
Surprisingly, the contrast of the phase space density between regions
inside and outside the caustics is much stronger in the data than in
both the $\tau$CDM and $\Lambda$CDM simulated clusters in D99.  The
difference may arise from the cosmological model used or the
semi-analytic techniques for defining galaxy formation and evolution
in the simulations.  The difference may be accentuated by the large
numbers of redshifts in the CAIRNS catalogs which extend
(non-uniformly) to fainter magnitudes than the simulated catalogs
displayed in D99.  The definition of the caustics is unlikely to be a
precise cosmological indicator, but it is suggestive that real
clusters more closely resemble the $\Lambda$CDM than the $\tau$CDM
simulated cluster.

The caustic pattern in the CAIRNS clusters is robust to the addition
of fainter galaxies to the radius-redshift diagrams \citep[see
also][]{gdk99}.  This result suggests that dwarf galaxies trace the
same caustic pattern as giant galaxies.

The D99 algorithm we use to identify the caustics does not always
agree with the lines one might draw based on a visual impression.  For
instance, A496 contains many galaxies 6-9 $\Mpc$ from the center and
at slightly lower redshift than the main body of the cluster.
However, the D99 method finds no significant caustic amplitude beyond
4 $\Mpc$.  One might expect the clump to indicate a large group or a
poor cluster, but these galaxies show no strong concentration on the
sky.  We conclude that this apparent structure is not part of A496 but
indicative of the surrounding large-scale structure.  A2199 also
contains many galaxies at the cluster redshift beyond the cutoff
radius of the caustics as well as extensive substructure including
X-ray groups. We discuss this system in detail in \citet{rines02}.
The outskirts of A539 contain a number of galaxies at the cluster
redshift but beyond the caustic cutoff radius.  The caustic signal
vanishes in the large radial gap at 3 $\Mpc$; the restriction on
steepness of the caustics prevents the amplitude from rising again.
Finally, the upper caustic of A168 encloses a relatively empty region
of phase space.  We discuss the individual clusters in more detail in
$\S 7.3$.

\subsection{Virial and Turnaround Radii}

The caustic mass profiles allow direct estimates of the virial and
turnaround radius in each cluster.  For the virial radius, we
calculate $r_{200}$, the density within which the enclosed average
mass density is 200 times the critical density $\rho _c$.  In our
adopted cosmology, a system should be virialized inside the slightly
larger radius $\sim$$r_{100} \approx 1.3 r_{200}$ \citep{ecf96}.  We
use $r_{200}$ because it is more commonly used in the literature and
thus allows easier comparison of results.  For the turnaround radius
$r_{t}$, we use equation (8) of \citet{rg89} assuming $\Omega _m =
0.3$.  For this value of $\Omega _m$, the enclosed density is 3.5$\rho
_c$ at the turnaround radius.  If the $w$ parameter in the equation of
state of the dark energy ($P_\Lambda = w\rho_\Lambda$) satisfies $w\ge
-1$, the dark energy has little effect on the turnaround overdensity
\citep[][]{2002MNRAS.337.1417G}.  Varying $\Omega_m$ in the range
0.02--1 only changes the inferred value of $r_t$ by $\pm$10\%; the
uncertainties in $r_t$ from the uncertainties in the mass profile are
comparable or larger \citep{rines02}.  

Table \ref{radii} lists $r_{200}$, $r_t$, and the masses $M_{200}$ and
$M_t$ enclosed within these radii.  The mass of the infall region is
20--120\% of the virial mass, demonstrating that clusters are still
forming in the present epoch.  Simulations of the future structure
formation of the nearby universe \citep{2002MNRAS.337.1417G,nl02} for
our assumed cosmology ($\Omega _m = 0.3, \Omega _\Lambda = 0.7$)
suggest that galaxies currently inside the turnaround radius of a
system will continue to be bound to that system.  In open cosmologies
with $\Omega_\Lambda = 0$, objects within which the enclosed density
exceeds the critical density are bound, whereas in closed cosmologies,
all objects are bound to all other objects.

One striking result of this analysis is that the caustic pattern is
often visible beyond the turnaround radius of a cluster.  This result
suggests that clusters may have strong dynamic effects on surrounding
large-scale structure beyond the turnaround radius.  For our assumed
cosmology, this large-scale structure is likely not bound to the
cluster. 

\subsection{The Ensemble CAIRNS Cluster}

Following other authors \citep[e.g.,][and references
therein]{cye97,bg03}, we construct an ensemble CAIRNS cluster to
smooth over the asymmetries in the individual clusters.  We construct
two versions of the ensemble cluster, one including all 9 clusters and
one excluding Coma, the best sampled cluster.  This procedure ensures
that our results are not biased by the inclusion of faint galaxies
which are better sampled in Coma than in the other clusters.  For both
cases, we scale the velocities by $\sigma_p$ (Table \ref{catalogs})
and positions with the values of $r_{200}$ determined from the caustic
mass profiles (Table \ref{radii}).

Figure \ref{combocaustics} shows the caustic diagrams for the ensemble
cluster including Coma.  The ensemble cluster excluding Coma yields
similar results, indicating that the results are robust to the
addition of faint galaxies \citep[see also][]{gdk99}.  The ensemble
caustic mass profile yields similar results to the individual
clusters.  At $r_{200}$, the mass is $M_{200} = (2.4\pm0.6)
\sigma_p^2 r_{200}/G$, consistent with the theoretical expectation of
$M_{200} = 3\sigma_p^2 r_{200}/G$.  The small difference suggests that
either the measured velocity dispersions are larger than the true
values (perhaps due to infalling, non-relaxed galaxies) or that the
estimates of $r_{200}$ are low.  

The ensemble cluster contains 3907 members within the caustics; 1746
of these are projected within $r_{200}$ and 1624 have projected radii
between 1 and 5 $r_{200}$ (5$r_{200}$ is comparable to the turnaround
radii in Table \ref{radii}).  These results show that the number of
galaxies in the infall region surrounding a cluster is comparable to
the number projected within the virial region.  Because the redshift
surveys are more deeply sampled in the cluster centers, the infall
region probably contains many more galaxies.  The ensemble CAIRNS
cluster contains more galaxies both within and outside $r_{200}$ than
either the CNOC1 ensemble cluster or that of \citet{bg03}.

\section{Comparison to Simple Parametric Models}

We fit the mass profiles of the CAIRNS clusters to three simple
analytic models.  The simplest model of a self-gravitating system is a
singular isothermal sphere (SIS). The mass of the SIS increases
linearly with radius.  \citet{nfw97} and \citet{hernquist1990} propose
two-parameter models based on CDM simulations of haloes.  We note that
the caustic mass profiles mostly sample large radii and are therefore
not very sensitive to the inner slope of the mass profile.  Thus, we
do not consider alternative models which differ only in the inner
slope of the density profile \citep[e.g.,][]{moore99}.  At large
radii, the NFW mass profile increases as ln$(r)$ and the mass of the
Hernquist model converges.  The NFW mass profile is
\beqn
M(<r) = \frac{M(a)}{\mbox{ln}(2) - \frac{1}{2}}[\mbox{ln}(1+\frac{r}{a})-\frac{r}{a+r}]
\eeqn
where $a$ is the scale radius and $M(a)$ is the mass within $a$. We
fit the parameter $M(a)$ rather than the characteristic density
$\delta_c$ (${M(a) = 4\pi \delta_c \rho_c a^3 [\mbox{ln}(2) -
\frac{1}{2}]}$ where $\rho_c$ is the critical density) because $M(a)$
and $a$ are much less correlated than $\delta_c$ and $a$
\citep{mahdavi99}.  The Hernquist mass profile is
\beqn
M(<r) = M \frac{r^2}{(r+a)^2}
\eeqn
where $a$ is the scale radius and $M$ is the total mass. Note that
$M(a) = M/4$. The SIS mass profile is 
\beqn
M(<r) = M(a=0.5~\Mpc) \frac{r}{0.5~\Mpc}
\eeqn
where we arbitrarily set the scale radius $a=0.5~\Mpc$.  We minimize
$\chi ^2$ and list the best-fit parameters $a$ (fixed for SIS) and
$M(a)$ for the three models  in Table \ref{mpfitsci}.  We perform the
fits on all data points within the turnaround radii listed in Table
\ref{radii} and with caustic amplitude $\mathcal{A} \mathnormal{^2(r)} >
100~\kms$.  The caustic amplitude becomes negligible at a radius
smaller than $r_t$ in 4 of the 9 clusters.  

Because the individual points in the mass profile are not independent,
the absolute values of $\chi ^2$ (Table \ref{mpfitsci}) are indicative
only, but it is clear that the NFW and Hernquist profiles provide
acceptable fits to the caustic mass profiles; the SIS is excluded for
all clusters.  The NFW profile provides a better fit to the data than
the Hernquist profile for 4 of the 8 CAIRNS clusters plus A194; the
other 4 are better fit by a Hernquist profile.  A non-singular
isothermal sphere mass profile yields results similar to the SIS;
thus, we report only our results for the SIS.  The concentration
parameters $c=r_{200}/a$ for the NFW models are in the range 5--17, in
good agreement with the predictions of numerical simulations
\citep{nfw97}.  There is no obvious correlation of $c$ with mass, but
the differences in $c$ should be small ($\sim$20\% and with large
scatter) over our mass range.

\section{Velocity Dispersion Profiles}

Several authors have explored the use of the velocity dispersion
profile (VDP) of clusters as a diagnostic of their dynamical states.
For example, \citet{1996ApJ...473..670F} find that VDPs typically have
three shapes: increasing, flat, or decreasing with radius.  

We calculate the VDPs of the CAIRNS clusters using all galaxies inside
the caustics.  All the CAIRNS clusters display decreasing VDPs within
about $r_{200}$ (Figure \ref{cairnsvdps}).  The VDPs either flatten
out or continue to decrease between $r_{200}$ and $r_t$.  Figure
\ref{cairnsvdps} also displays the predicted VDPs of the best-fit
Hernquist mass models calculated assuming isotropic orbits.  Because
these predicted VDPs have no free parameters, they provide an
interesting consistency check on our caustic mass profiles.  The
observed VDPs agree with the models for most clusters, even at radii
much larger than $r_{200}$.  The most significant difference is found
in A2199, where the predicted VDP is smaller than the observed VDP.
The observed VDP might be artificially enhanced by groups within
the infall region \citep{rines01b,rines02}, or the caustic mass
profile might be too small.  Indeed, using the most massive profile
from a more detailed study of this system \citep{rines02} yields a
predicted VDP that matches the observed VDP.  Figure \ref{combovdp}
shows that the predicted and observed VDPs of the ensemble cluster are
in good agreement.  This agreement suggests that the caustic mass
profiles are consistent with Jeans analysis (see $\S 7.2$) and that
the orbits are not far from isotropic, even outside the virial radius.
However, this result could perhaps be mimicked by a population of
infalling galaxies not in equilibrium and not on radial orbits.
Indeed, many authors find evidence of an infalling population
dominated by blue or emission line galaxies with larger velocity
dispersions than the red and presumably more relaxed galaxies
\citep{mohr96,1997ApJ...476L...7C,mahdavi99,kg2000,ellingson01}.

Some authors \citep{cye97,girardi98} claim that an accurate estimate
of $r_{200}$ can be obtained for a cluster from the asymptotic
value of the velocity dispersion calculated for all galaxies within a
given radius.  Many of the CAIRNS clusters, however, display no
obvious convergence in the enclosed velocity dispersion (shown by
triangles in Figure \ref{cairnsvdps}).  If the rich clusters in the
CNOC1 survey are similar to their CAIRNS cousins, the use of the
asymptotic value of the enclosed velocity dispersion to estimate
$r_{200}$ may be unreliable.  The caustic technique provides an
alternative method for estimating $r_{200}$, although applying it
requires many more redshifts than are needed for computing the
velocity dispersion.

\section{Comparison to Other Mass Estimates}

Because the caustic mass estimator is calibrated by simulations, it is
important to test the accuracy of the estimator with independent
observational mass estimates.  Clusters usually possess a hot
intracluster medium (ICM) with temperatures of $10^7$--$10^8$ K.
Assuming this hot gas is in hydrostatic equilibrium, one can estimate
both the mass contained in hot gas and the total gravitational mass in
the X-ray emitting region \citep{flg80}.  The caustic method is
completely independent of the X-ray properties of the clusters.

\subsection{Mass and X-ray Temperature}

The mass-temperature relation
\citep{emn96,1999ApJ...520...78H,2000ApJ...532..694N, frb2001} gives a
straightforward estimate of the mass of a cluster from its X-ray
temperature.  We use the mass-temperature relation rather than, e.g.,
a $\beta$ model to estimate the mass both for simplicity and to ensure
uniformity (X-ray observations allowing more detailed mass estimates
are not available for all CAIRNS clusters).  Numerical simulations
\citep{emn96} suggest that estimating cluster masses based solely on
emission-weighted cluster temperatures yields similar accuracy and
less scatter than estimates which incorporate density information from
the surface brightness profile.  In $\S 7.3$, we compare our results
to more detailed X-ray mass estimates in the literature.

\citet{frb2001} find a relation of $M_{500} = (1.87\pm 0.14) \times
10^{13} T_{keV}^{1.64\pm0.04} \msun$, where $M_{500}$ is the enclosed
mass within the radius $r_{500}$ and $T_{keV}$ is the
emission-weighted electron temperature in keV.  The X-ray temperature
yields an estimate for $r_{500}$ of $r_{500} = (0.32\pm0.01)
\sqrt{T_{keV}}~\Mpc$.  The temperatures of the four clusters in both
their sample and ours are consistent.  Figure \ref{cx} displays the
interpolated caustic mass estimate at $r_{500}$ (we derive $r_{500}$
from $T_{keV}$ rather than directly from the caustic mass profile)
versus $M_{500}$ from the mass-temperature relation.  The agreement is
generally good, with the exception of A576.  The ratio of the caustic
estimate to the X-ray estimate has an error-weighted mean of $1.03\pm
0.11$, indicating that the two methods are in excellent agreement.  If
we eliminate A576, an outlier in Figure \ref{cx}, the error-weighted
mean ratio is $0.97\pm 0.11$ \citep[][discusses A576 in more
detail]{rines2000}.  The scatter in the mass estimates of individual
clusters is therefore $\sim$33\%.  The excellent agreement between the
caustic technique and X-ray mass estimates confirms the prediction of
D99 that the caustic mass estimate is unbiased.  The relatively small
scatter in the ratio suggests that the systematic uncertainty in the
caustic technique is actually smaller than the factor of two suggested
by simulations (see $\S 4.1$ and D99).

\subsection{Virial and Projected Mass Estimates}

\citet{zwicky1933,zwicky1937} used the virial theorem to estimate the
mass of the Coma cluster.  With some modifications, notably a
correction term for the surface pressure \citep{1986AJ.....92.1248T},
the virial theorem remains in wide use \citep[e.g.,][and references
therein]{girardi98}.  Jeans analysis incorporates the radial
dependence of the projected velocity dispersion \citep[e.g.,][and
references therein]{cye97,2000AJ....119.2038V,bg03} and obviates the
need for a surface term.

Jeans analysis and the caustic method are closely related.  Both use
the phase space distribution of galaxies to estimate the cluster mass
profile.  The primary difference is that the Jeans method assumes that
the cluster is in dynamical equilibrium; the caustic method does not.
The Jeans method depends on the width of the velocity distribution of
cluster members at a given radius, whereas the caustic method
calculates the edges of the velocity distribution at a given radius.
The caustic method is not independent of the Jeans method, as the D99
method generally requires $<v_{esc}^2>_R \sim 4 <v^2>_R$ within the
virial region with radius R (see D99 for a more detailed
discussion).  Mass estimates based on Jeans analysis
thus provide a consistency check but not an independent verification
of the caustic mass estimates.

Applying the Jeans method requires an assumption about either the mass
distribution or the orbital distribution.  Typically, one assumes that
light traces mass and thus that the projected galaxy density is
proportional to the projected mass density \citep[e.g.,][]{girardi98}
or one assumes a functional form for the orbital distribution
\citep[e.g.,][]{bg03}.  Note that most authors make the implicit
assumption that the orbital distribution of the dark matter can be
inferred from that of the galaxies. \citet{dkthesis} shows that
galaxies in simulated clusters often have significantly different
orbital distributions than the dark matter.  Unfortunately, our
redshift catalogs are highly non-uniform due to the use of data from
the literature and due to the systematic uncertainties inherent in
photographic photometry on large scales.  This non-uniformity prevents
an estimate of the projected galaxy density to a consistent magnitude
limit.

We apply the virial mass and projected mass estimators
\citep{htb} to the CAIRNS clusters.  For the latter, we assume the
galaxies are on isotropic orbits.  We must define a radius of
virialization within which the galaxies are relaxed.  We use $r_{200}$
(Table \ref{radii}) and include only galaxies within the caustics.  We
thus assume that the caustics provide a good division between cluster
galaxies and interlopers (see Figure \ref{allcaustics}).  We test the
Jeans analysis on Coma by assuming a Hernquist profile and the scale
radius from the caustic mass profile; we fit a constant anisotropy
$\beta$ and the core mass $M(a)$.  The best-fit core mass is $39\pm5
\times 10^{13} \msun$ (68\% confidence level), in excellent agreement
with the core mass from the caustics.  The best-fit anisotropy is
$\beta = -0.4^{+0.6}_{-1.0}$, slightly tangentially anisotropic but
consistent with isotropic orbits.  For the ensemble cluster, the
best-fit anisotropy is $\beta = -0.05^{+0.22}_{-0.27}$, again
consistent with isotropic orbits.  The best-fit core mass again agrees
very well with the caustic mass profile.  For this analysis, we
calculate the VDPs from all galaxies within the caustics and within
$r_{200}$.  The caustics contain infalling galaxies which are not in
equilibrium.  Including these galaxies might result in an overestimate
of the mass.

We calculate the virial mass according to
\begin{equation}
M_{vir} = \frac{3 \pi}{2} \frac{\sigma_p^2 R_{PV}}{G}
\end{equation}
where $R_{PV} =  2N(N-1)/\sum_{i,j>i}R_{ij}^{-1}$ is the projected
virial radius and $\sigma_p^2 = \sum_i (v_i-\bar v)^2/(N-1)$. 
If the system does not lie entirely within $r_{200}$, a surface
pressure term 3PV should be added to the usual virial theorem so that
$2T + U = 3PV$. The virial mass is then an overestimate of the mass
within $r_{200}$ by the fractional amount 
\begin{equation}
\label{virialc}
C =  4\pi r_{200}^3 \frac{\rho (r_{200})}{\int_{0}^{r_{200}}4\pi r^2 \rho dr} \Bigg[{\frac{\sigma _{r} (r_{200})}{\sigma (<r_{200})}\Bigg]^2}
\end{equation}
where $\sigma_r (r_{200})$ is the radial velocity dispersion at
$r_{200}$ and $\sigma (<r_{200})$ is the enclosed total velocity
dispersion within $r_{200}$ \citep[e.g.,][]{girardi98}.  In the
limiting cases of circular, isotropic, and radial orbits, the maximum
value of the term involving the velocity dispersion is 0, 1/3, and 1
respectively. 

The projected mass estimator is more robust in the presence of close
pairs.  The projected mass is
\begin{equation}
M_{proj} = \frac{32}{\pi G} \sum_i{R_i (v_i-v)^2}/N
\end{equation}
where we assume isotropic orbits and a continuous mass distribution.
If the orbits are purely radial or purely circular, the factor 32
becomes 64 or 16 respectively.  We estimate the uncertainties using
the limiting fractional uncertainties $\pi^{-1} (2 \mbox{ln}
N)^{1/2}N^{-1/2}$ for the virial theorem and $\approx 1.4 N^{-1/2}$
for the projected mass.  These uncertainties do not include systematic
uncertainties due to membership determination or the assumption of
isotropic orbits in the projected mass estimator.
Table \ref{virial} lists the virial and projected mass estimates. 

Figure \ref{vp} compares the projected mass estimates to the virial
mass estimates.  The two mass estimators generally give consistent
results, but the uncertainties are probably underestimated.  The
error-weighted mean ratio of the estimates is $M_p/M_v = 1.18 \pm
0.05$.  This result is not dominated by the outlier, A539.  If we
eliminate this data point, the mean ratio is $1.16\pm 0.05$.
Similarly, Figures \ref{vc} and \ref{pc} compare these two estimators
to the caustic mass estimates at $r_{200}$.  The error-weighted mean
ratios of these estimates are $M_c/M_v = 0.79 \pm 0.05$ and $M_c/M_p =
0.68 \pm 0.05$.  The caustic mass estimates are consistent with virial
mass estimates assuming a correction factor $C\approx 0.2 M_{vir}$,
consistent with the best-fit NFW profiles and in good agreement with
\citet{cye97}, who estimate the correction needed for orbital
distributions with constant $\beta$ in the range
-0.50$\leq\beta\leq$+0.75.  The squared velocity dispersion term in
Equation \ref{virialc} is $\sim$0.10-0.33 for isotropic orbits (This
term is a factor of 3 larger for purely radial orbits).  The rest of
the right-hand side of Equation \ref{virialc} varies from $\sim$0.72
for $c$=5 to $\sim$0.40 for $c$=25.  Indeed, a crude estimate of $C$
from these parameters (Table \ref{virial}) yields excellent agreement
with the caustic mass estimates, with an error-weighted mean of
$M_c/M_{cv}$ = $0.93 \pm 0.07$ where $M_{cv} = (1-C)M_v$ (Figure
\ref{cvc}).  The correction factor is $C$=0.05-0.22, but this
correction factor could be larger by a factor of 3 for purely radial
orbits.  Such radially anisotropic orbits would, however, produce more
steeply decreasing VDPs \citep[see][]{girardi98}.  This comparison
confirms the indications from the VDPs ($\S 6$) that the caustic
technique yields mass estimates consistent with Jeans analysis.

In their ensemble cluster, \citet{bg03} found that the caustic mass at
$r_{200}$ is $\sim$60\% of the mass determined from Jeans analysis
(restricted to galaxies with absorption-dominated spectra and assuming
isotropic orbits) when they use a fixed value of $F_\beta (r) = 0.5$
to determine the caustic mass.  If they use a non-constant $F_\beta
(r)$ with a shape similar to that in Figure 3 of D99, the two mass
estimates agree well.  However, as shown above, the assumption of
$F_\beta (r) = 0.5$ for the CAIRNS clusters yields caustic mass
estimates consistent with Jeans analysis and in excellent agreement
with the virial theorem if the surface pressure term is included.
Also, using a non-constant $F_\beta (r)$ requires knowledge of
$r_{200}$, whereas assuming a constant value does not.  The caustics
of the ensemble 2dF cluster appear to be flatter (less centrally
peaked) than those of CAIRNS clusters; this difference, which perhaps
reflects the difference in the typical cluster mass (CAIRNS clusters
are generally more massive), may account for the different
conclusions.  Alternatively, aperture bias may cause \citet{bg03} to
include misclassified emission-dominated galaxies not in equilibrium
in the Jeans analysis, thus biasing the Jeans mass toward larger
values.  However, we see no such bias in the above Jeans analysis of
CAIRNS clusters despite making no attempt to remove infalling,
non-equilibrium galaxies.  Other possible differences include the
deeper sampling of CAIRNS clusters and the relatively larger
uncertainties in the scaling parameters in \citet{bg03}.

\subsection{Individual Clusters}
 
Previous studies have shown good agreement between caustic mass
profiles and X-ray and virial estimates in individual systems: Coma
\citep{gdk99}, A576 \citep{rines2000}, Fornax \citep{drink}, A1644
\citep{tustin}, the Shapley Supercluster (RQCM), and A2199
\citep{rines02}.  We discuss the five other CAIRNS clusters and A194 
individually below.

\subsubsection{Abell 119}

In a previous study, \citet{1993AJ....105..788F} demonstrate that A119
has a significant extension to the NE both in projected surface galaxy
density and in X-ray emission.  The redshift survey confirms that the
overdensity of galaxies is at the same redshift as A119 and that it
extends significantly farther to the NE than the study of \citet[][
see Rines et al.~in preparation]{1993AJ....105..788F}.  Figure
\ref{a119mp} shows the mass profile of A119.  The caustic mass profile
agrees very well with both X-ray and virial mass estimates compiled by
\citet{girardi98}.  There are many other determinations of the mass of
A119 from X-ray data, and they give a range of results.  This range
may result from calibration uncertainties in the various satellites:
{\em Einstein} \citep{wjf97,jf99}, {\em ROSAT} and {\em ASCA}
\citep{mme,frb2001,hiflugcs}, {\em ROSAT} and {\em Einstein}
\citep{cirimele97,peres98}, and {\em BeppoSAX} \citep{ettori02}.

Because the virial masses calculated in $\S 7.2$ are based on the
largest redshift samples currently available, we do not plot previous
determinations of the virial masses in Figure \ref{a119mp} or similar
figures in the following subsections.

\subsubsection{Abell 168}

\citet{1992ApJ...397..430U} studied the significance of X-ray-galaxy
count offsets in clusters and concluded that A168 showed significant
evidence of an offset, suggesting an ongoing merger between two
roughly equal mass groups with different gas-to-galaxy ratios.  We
confirm this X-ray-optical offset in Table \ref{centers}, where we
find a separation of 239$~\kpc$ between the X-ray center and the
hierarchical center of the galaxy distribution.
\citet{1996ApJ...473..670F} find that A168 contains three components
in redshift space.  \citet{girardi98} calculate the virial mass of
A168 both in a joined and disjoined analysis (the joined analysis
treats the system as a single body wherease the disjoined analysis
treats the different redshift components as multiple systems). The
joined peaks calculation yields $r_{vir}=0.87~\Mpc$ and
$M_{vir}=1.5\pm0.25 \times 10^{14} h^{-1} M_\odot$; the disjoined
peaks have masses smaller by an order of magnitude.  Figure
\ref{a168mp} shows the mass estimates of the joined peaks analysis and
the most massive of the three systems in the disjoined analysis.
Our mass estimates exceed all of these estimates. 

Figure \ref{a168mp} shows the mass estimated from the X-ray gas by
\citet{wjf97} and the estimate from the $M_{500}-T_X$ relation.  Both
estimates are smaller than the caustic mass at the same radii.

Our redshift survey shows that the offset between the peak of X-ray
emission and galaxies at the cluster redshift is a real, physical
offset (Rines et al.~in preparation).  Outside 1.7$~\Mpc$, the
caustics of A168 are only constrained by the restriction on the
derivative of $\mathcal{A} \mathnormal{(r)}$. This fact suggests that
the caustic technique may be an unreliable measure of the membership
in A168.  We have previously shown that the caustic technique produces
accurate estimates of mass profiles in minor mergers of well-separated
clusters like the A2199/A2197 supercluster.  The presence of an
ongoing major merger in the center of A168 may heat the gas above a
hydrostatic temperature; the X-ray temperature might thus be an
inaccurate mass estimator.  Alternatively, the caustic technique may
fail in the presence of an ongoing major merger.  Regardless of these
issues, the difference is no larger than about a factor of two, well
within the range of projection effects.

\subsubsection{Abell 496}

A496 has very symmetric X-ray emission. \citet{mvfs} suggest that A496
and A2199 are prototypical relaxed clusters ideal for estimating mass
profiles based on their symmetry, temperature profiles, and evidence
for moderate cooling flows.  \citet{durret00} analyze both X-ray data
and a large catalog of redshifts and concur that A496 seems to be a
relaxed cluster in both X-rays and in the galaxy distribution, with
the possible exception of emission line galaxies.  Like A119, there
are many X-ray mass estimates of A496, all of which are shown in
Figure \ref{a496mp}.  These estimates use data from various
satellites: {\em Einstein} \citep{durret94,wjf97,jf99}, {\em ROSAT}
\citep{durret00}, {\em ROSAT} and {\em ASCA} 
\citep{mvfs,mme,frb2001,hiflugcs}, {\em ROSAT} and {\em Einstein}
\citep{peres98}, and {\em BeppoSAX} \citep{ettori02}.  The
scatter in the X-ray estimates is significantly smaller than in A119,
and the agreement with the caustic mass profile is excellent.  We
found similar agreement in A2199 \citep{rines02}, suggesting that
X-ray emission from relaxed clusters yields very robust mass
estimates.

\subsubsection{Abell 539}

In a study of A539, \citet{ostriker88} first looked for the caustic
pattern around a cluster.  Despite measuring several hundred
redshifts, they were unable to place good constraints on the infall
properties of A539.  Even with the much larger redshift sample here,
there are relatively few galaxies in the infall region of A539 (see
Figure \ref{allcaustics} and Table \ref{radii}).  The virial mass and
projected mass estimator yield significantly different estimates of
the mass of A539 (see Table \ref{virial}), the largest difference of
any of the CAIRNS clusters (Figure \ref{vp}).  The virial mass
estimate of \citet{girardi98} is intermediate between our virial mass
and projected mass estimates.

\citet{djf95,djf96} analyzed {\em ROSAT} X-ray observations of A539
and found an X-ray temperature of $1.3^{+0.8}_{-0.3}$~keV, unusually
cool for a rich cluster.  Assuming a hydrostatic-isothermal $\beta$
model, \citet{djf95} found a cluster mass of $(1.0\pm0.5)\times
10^{14}\msun$, significantly smaller than virial mass estimates.
\citet{djf96} are unable to resolve this discrepancy using simple
models for the orbital distribution of the galaxies.  X-ray
observations of A539 using other instruments yield higher ICM
temperatures, i.e., $3.0^{+0.5}_{-0.4}$~keV for {\em Einstein} MPC
data \citep{david93} and $3.2\pm0.1$~keV \citep{vfj99} and
$3.7\pm0.3$~keV \citep{white2000} for ASCA data. These comparisons
suggest that the X-ray-optical mass discrepancy found in \citet{djf96}
is due to a large systematic error in the {\em ROSAT} temperature
determination.  We therefore recalculate the estimate of \citet{djf96}
using $T_X = 3.0$~keV and plot this estimate in Figure \ref{a539mp}
along with X-ray mass estimates from {\em Einstein} data
\citep{wjf97,jf99}, {\em ROSAT} and {\em ASCA}
\citep{frb2001,hiflugcs}, and the $M_{500}-T_X$ relation.  The X-ray
mass estimates agree reasonably well with the caustic technique.

\subsubsection{Abell 1367}

\citet{girardi98} find that A1367 has multiple peaks in the velocity
distribution. They calculate two masses for the main body of the
system by assuming that the peaks are either joined or disjoined
(i.e., one or two distinct bodies).  Figure \ref{a1367mp} shows the
mass estimates of the joined peaks and the more massive of the
disjoined peaks.  The caustic mass profile is closer to the mass of
the joined peaks but is consistent with the estimates of both the
joined and disjoined peaks.  The caustic technique is sensitive to the
total interior mass and is thus more similar to the joined peaks
analysis.  The agreement of the two techniques suggests that the
caustic technique is reasonably accurate even in the presence of
significant velocity substructure in the center of a cluster.

\citet{hanka1367} use {\em ROSAT} and {\em ASCA} X-ray observations to 
show that A1367 is undergoing a merger of two subclusters. The X-ray
image shows two peaks separated by $11\farcm5=0.21\Mpc$ on the
sky. The optical center we find from a hierarchical cluster analysis
is located $4\arcm$ N of the more X-ray luminous SE peak and $8\arcm$
SE of the less luminous NW peak.  The X-ray temperature varies
significantly across the system and suggests that the ICM is shock
heated, at least in the less luminous but hotter NW subcluster.  The
additional heating due to shocks and the presence of multiple peaks
complicate the mass estimate.  \citet{hanka1367} model the system as
two distinct clusters and calculate masses for each subcluster at
radii of 0.25 and $0.50~\Mpc$.  Because the optical center is located
between the X-ray peaks, we sum the masses of the subclusters to
estimate the total mass of the system (the caustic technique should
give the combined mass of all bodies in the system).  Figure
\ref{a1367mp} shows the mass estimates derived using the observed
X-ray temperature and gas distribution, but note that
\citet{hanka1367} suggest that the mass might be lower by as much as
$\sim$40\% if non-thermal heating is significant.  \citet{white2000}
finds a flat temperature profile in A1367 consistent with an
isothermal temperature of $4.0\pm0.5$keV, consistent with the
temperature in Table \ref{sample}.

Several authors have estimated the mass of A1367 without explicit
discussion of substructure; we assume that these estimates are for the
more luminous SE component.  These estimates use data from
various satellites: {\em Einstein} \citep{wjf97,jf99},  {\em
ROSAT} and {\em ASCA} \citep{mme}, and {\em ROSAT} and {\em Einstein}
\citep{peres98}.  Figure \ref{a1367mp} shows these estimates along
with mass at $r_{500}$ estimated using the $M_{500}-T_X$ relation.

The complexities described by \citet{hanka1367} suggest that X-ray
mass estimates for A1367 could contain significant systematic
uncertainties.  Similarly, the velocity distribution is more
complex than expected for a relaxed cluster \citep{girardi98}.
The caustic technique yields a mass profile contained within the
ranges of previous mass estimates based on optical and X-ray
data. This result suggests that the caustic technique is effective
even in complicated clusters containing significant substructure
\citep[see also][]{rqcm,rines01b,rines02}.

\subsubsection{Abell 194}

Abell 194 is a curious example of a cluster that is significantly
elongated in both X-rays and in its galaxy distribution
\citep[see][and references therein]{1988AJ.....95..999C}.
\citet{1988AJ.....95..999C} show that A194 is anisotropic and
investigate the effects of anisotropy on the virial mass estimator.
They conclude that the virial mass of A194 is $1.8^{+0.8}_{-0.4}
\times 10^{14} h^{-1} M_\odot$ where the asymmetric uncertainties
reflect analytic calculations of the uncertainties due to an
anisotropic galaxy distribution.  \citet{girardi98} estimate a virial
(corrected virial) mass of $7.0^{+2.5}_{-1.8} (6.0^{+2.2}_{-1.5})
\times 10^{13} h^{-1} M_\odot$ at $0.68~\Mpc$.  These estimates are
consistent with our results.  Finally, we note that
\citet{1998AJ....116.1573B} detected two compact groups within A194,
although they conclude that only one of them is a physical system,
perhaps a cold core of galaxies in the center of the cluster
potential.

\citet{wjf97} find a mass of $2.5\pm1.1 \times 10^{13} h^{-1} M_\odot$
within $0.25~\Mpc$ from a deprojection analysis of {\em Einstein} data.
\citet{1999A&A...349...97N} analyze multiwavelength observations of
A194 and conclude that the significant elongation in the galaxy
distribution is present only in the center of the cluster.  They
estimate a mass of $4.0^{+6.0}_{-2.5} \times 10^{13} h^{-1} M_\odot$
within $0.4~\Mpc$ based on {\em ROSAT} PSPC data.  Figure \ref{a194mp}
displays the caustic mass profile of A194, the X-ray and optical
estimates from $\S\S 7.1$ and 7.2, and the X-ray estimates from the
literature.   The caustic mass profile agrees well with all other
estimates, although the uncertainties are large.

\section{Discussion}

We catalog \ncztot redshifts in the infall regions of eight rich,
nearby, X-ray luminous clusters in CAIRNS along with A194, a ``bonus''
cluster in the foreground of one of the eight CAIRNS clusters.  CAIRNS
is the largest redshift survey of cluster infall regions to date.  We
find the following:
\begin{itemize}
\item{
The expected caustic pattern in redshift space
is present in all systems.  }
\item{ We use the shape of the caustics to
calculate the mass profiles of the systems.  These mass profiles yield
estimates of the virial radii and the turnaround radii.  The caustic
pattern is often visible up to and beyond the turnaround radius.  The
mass in the infall region is 20--120\% of the virial mass, showing that
clusters are still forming.  }
\item{We stack the clusters to produce an ensemble cluster containing
1746 galaxies projected within $r_{200}$ and an additional 1624 within
5$r_{200}$ (roughly the turnaround radius). The infall region thus
contains at least as many galaxies as the virial region.}
\item{
The caustic mass profiles agree with Hernquist
and NFW models but exclude an isothermal sphere.  Observed clusters
resemble those in simulations, and their mass profiles are well
described by extrapolations of NFW or Hernquist models out to the
turnaround radius.  }
\item{ The velocity dispersion profiles of the best-fit Hernquist mass
profiles agree with the observed VDPs assuming isotropic orbits. }
\item{ The observed VDPs all decrease with radius within $r_{200}$.}
\item{ At small radii, the mass
profiles agree with independent X-ray mass estimates with a mean ratio
of $1.03\pm 0.11$.}
\item{ The scatter in the ratio of the caustic mass estimate to X-ray
estimates in observations ($\sim$33\%) appears to be smaller than the
scatter in the ratio of the caustic mass estimate to the true mass
profile in simulations ($\sim$50\%). }
\item{ At larger
radii, the caustic masses are smaller than projected masses and virial
masses.  However, the caustic masses agree with virial masses when the
surface pressure term is taken into account (the mean ratio is
$0.93\pm0.07$).  The caustic masses agree with masses
from Jeans analysis (assuming a Hernquist model) for both Coma and the
ensemble cluster.  These fits also indicate that the orbital
distribution is close to isotropic. }
\item{ These results and those in
the literature demonstrate the power of the caustic technique as a
mass estimator over a large dynamic range in both mass and radius,
from systems as small as the Fornax cluster \citep[$5-9 \times 10^{13}
M_\odot$;][]{drink} to the most massive ($1-2 \times 10^{15}
\msun$) clusters like Coma \citep[]{gdk99} and the Shapley
Supercluster \citep{rqcm}. }
\item{ The mass profiles of individual clusters
agree well with detailed X-ray mass estimates in the literature, and
it is striking that A496, a relaxed cluster, shows the best agreement
between caustic and X-ray mass estimates as well as the smallest
scatter in the X-ray estimates.  Two clusters, A168 and A1367, are
probably undergoing major mergers.  The caustic technique yields mass
estimates comparable to other estimators even in these complex
systems. }
\end{itemize}

The CAIRNS project demonstrates that the caustic pattern is ubiquitous
in rich, X-ray luminous galaxy clusters.  The large 2dF and Sloan
redshift surveys will provide large numbers of spectra which can be
used to explore the caustic effect in a greater number and greater
range of clusters.  For instance, \citet[][]{bg03} analyze the 2dF
100,000 redshift data release and produce an ensemble cluster of 43
clusters generally less massive than the CAIRNS clusters.  Because
only $\sim$1\% of galaxies reside in rich clusters, the CAIRNS survey,
which specifically targets clusters, actually yields more cluster
redshifts than the 2dF 100,000 redshift data release.  Future papers
in the CAIRNS project will analyze the relative distributions of mass
and light in cluster infall regions, X-ray and optical substructure
within infall regions, and the dependence of spectroscopic properties
on environment.

\acknowledgements

This project would not have been possible without the assistance of
Perry Berlind and Michael Calkins, the remote observers at FLWO, and
Susan Tokarz, who processed the spectroscopic data.  We thank Dan
Fabricant for orchestrating the design and construction of the FAST
spectrograph which has kept the modest 1.5-m Tillinghast telescope
surprisingly competitive.  KR, MJG, and MJK are supported in part by
the Smithsonian Institution.  The National Geographic Society -
Palomar Observatory Sky Atlas (POSS-I) was made by the California
Institute of Technology with grants from the National Geographic
Society.  This research has made use of the NASA/IPAC Extragalactic
Database (NED) which is operated by the Jet Propulsion Laboratory,
California Institute of Technology, under contract with the National
Aeronautics and Space Administration. We thank Marion Schmitz at IPAC
for pointing out errors in our online redshift catalog around Coma.
We thank Dan Fabricant, Bill Forman, Lars Hernquist, Christine Jones,
Andi Mahdavi, Joe Mohr, and Gary Wegner for helpful discussions.  We
thank Joe Mohr and Gary Wegner for the use of unpublished redshifts
near A576 and A2199.  We acknowledge the Max-Planck-Institut f\"ur
Astrophysik in Garching for use of its computer facilities for some
computations.  We thank the referee for comments which improved the
presentation of this paper.

\bibliographystyle{apj}
\bibliography{rines}

\clearpage

\begin{figure}
\figurenum{4}
\plotone{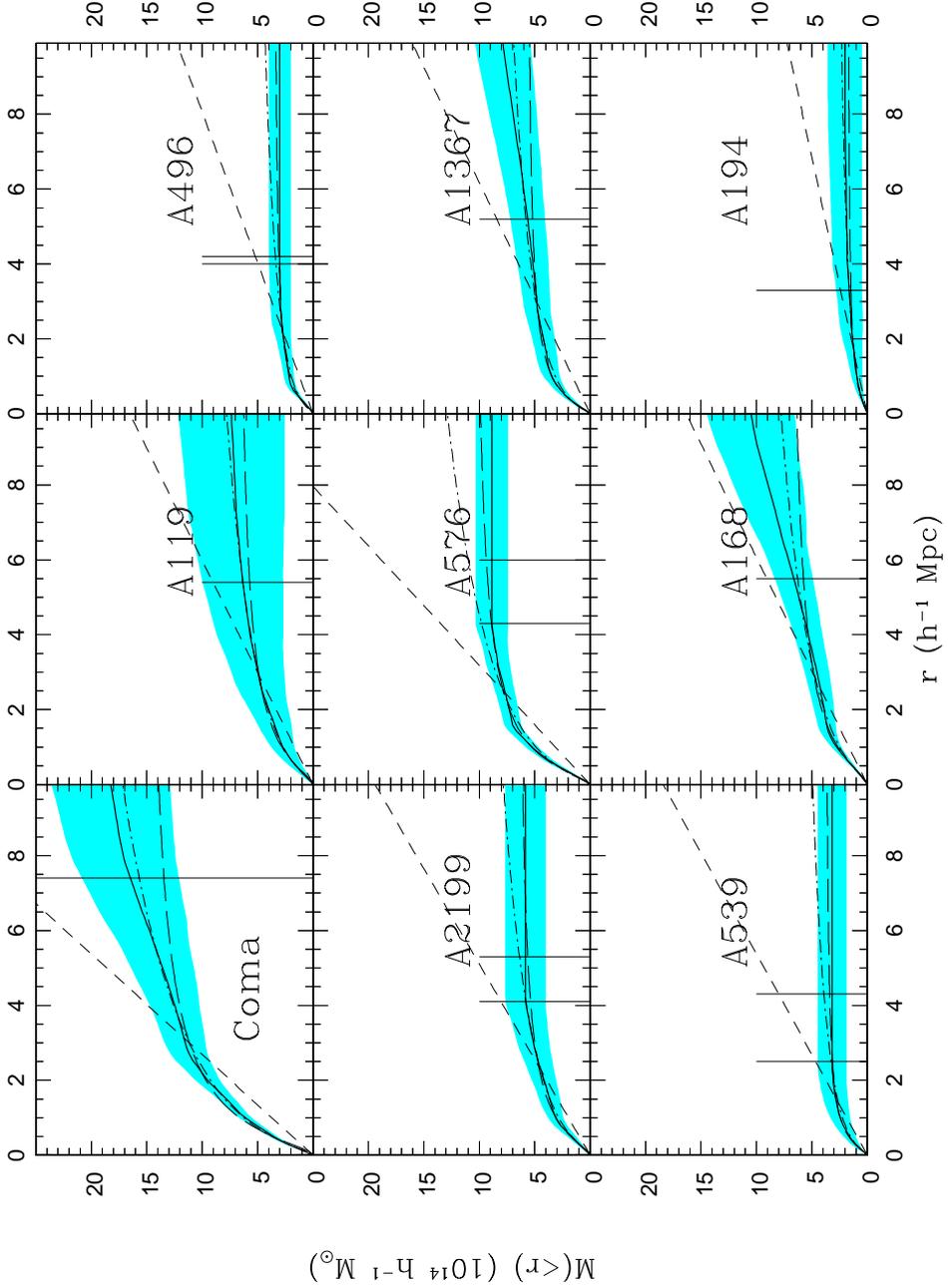} 
\caption{\label{allmp2} Caustic mass profiles for the CAIRNS
clusters compared to simple models. The thick solid lines show the
caustic mass profiles and the shaded region shows the 1-$\sigma$
uncertainties in the mass profiles.  The vertical bars indicate the
turnaround radius and the limit of the caustics if it is smaller than
the turnaround radius.  The best-fit SIS, NFW, and Hernquist profiles
are displayed as short-dashed, dash-dotted, and long-dashed lines
respectively.  The axes are identical in all panels.}
\end{figure}

\begin{figure}
\figurenum{5}
\plotone{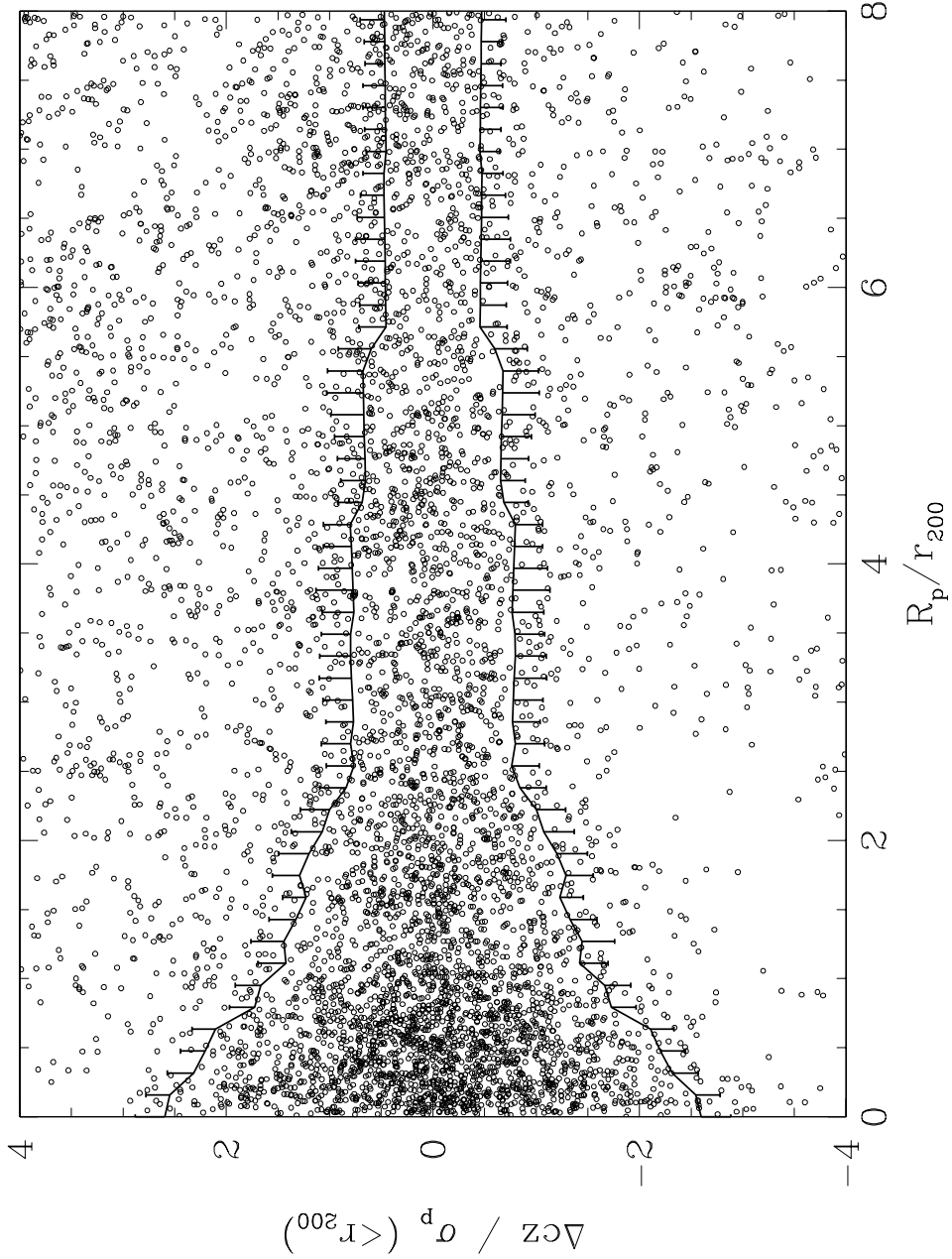} 
\caption{\label{combocaustics} Redshift versus radius for galaxies around
the CAIRNS ensemble cluster. The solid lines indicate our estimate of
the location of the caustics.  The errorbars are 1-$\sigma$
uncertainties and are shown only on one side of each caustic for
clarity. }
\end{figure}

\begin{figure}
\figurenum{6}
\plotone{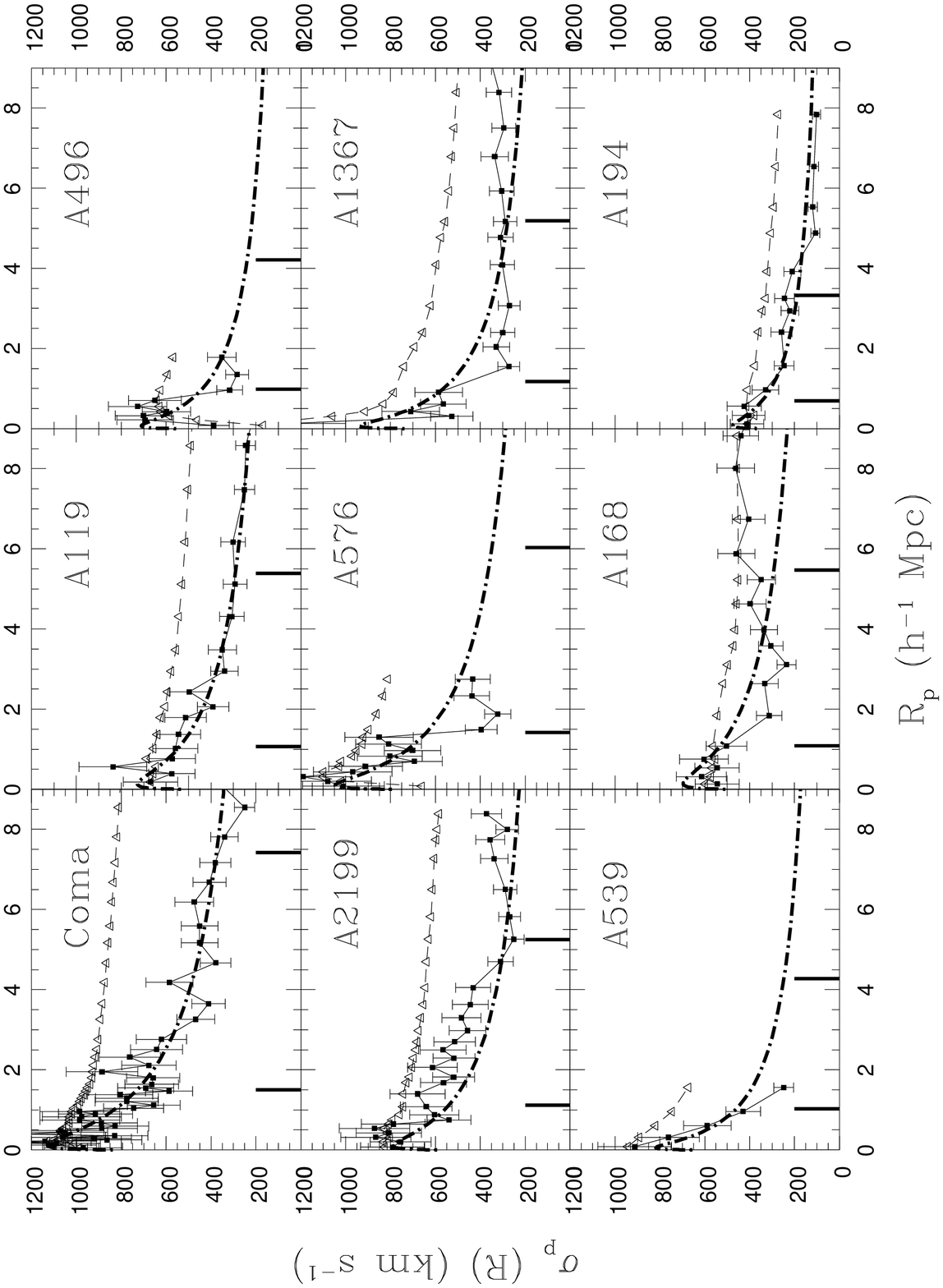} 
\caption{\label{cairnsvdps} Velocity dispersion profiles for the CAIRNS
clusters. The filled squares and solid lines show the velocity
dispersion profile of member galaxies (those within the caustics) with
1-$\sigma$ uncertainties.  The open triangles and dashed lines show
the enclosed velocity dispersion.  The dash-dotted lines show the VDPs
predicted by the Hernquist mass models which best fit the caustic mass
profiles (assuming isotropic orbits).  Heavy ticks on the abscissa
indicate $r_{200}$ and $r_t$. The axes are identical in all panels.}
\end{figure}

\begin{figure}
\figurenum{7}
\epsscale{1.0}
\plotone{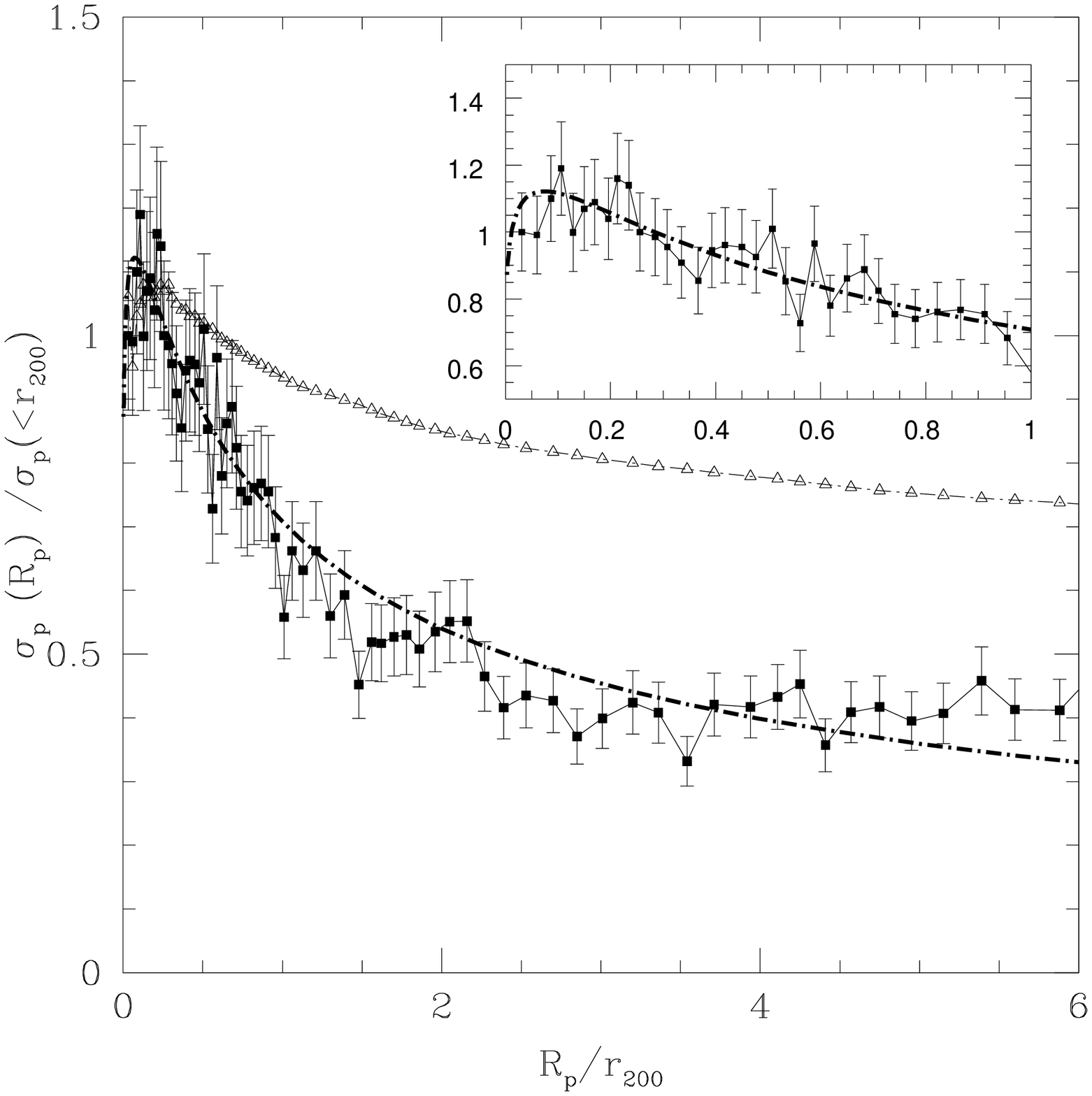} 
\caption{\label{combovdp} Velocity dispersion profile of the ensemble CAIRNS
cluster. The filled squares and solid line show the velocity
dispersion profile of member galaxies (those within the caustics) with
1-$\sigma$ uncertainties.  The open triangles and dashed line show the
enclosed velocity dispersion.  The dash-dotted line shows the VDP
predicted by the Hernquist mass models which best fits the caustic
mass profiles (assuming isotropic orbits).  The inset shows a closeup
of the VDP within $r_{200}$.}
\end{figure}

\begin{figure}
\figurenum{8}
\plotone{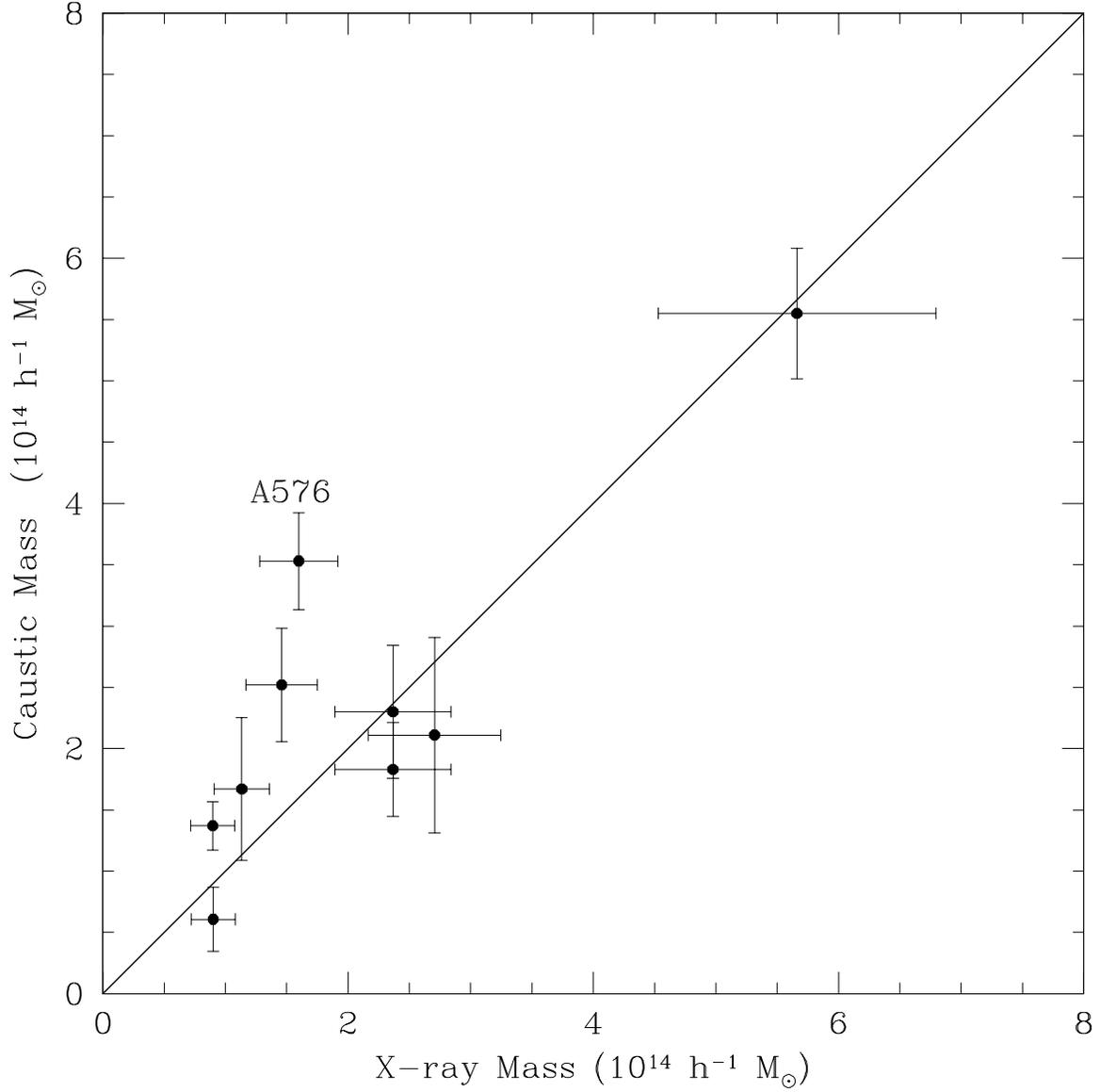} 
\caption{\label{cx} Comparison of caustic mass estimates to estimates
based on the mass-temperature relation.  The solid line has a slope of
unity and intercepts the origin.} 
\end{figure}

\begin{figure}
\figurenum{9}
\plotone{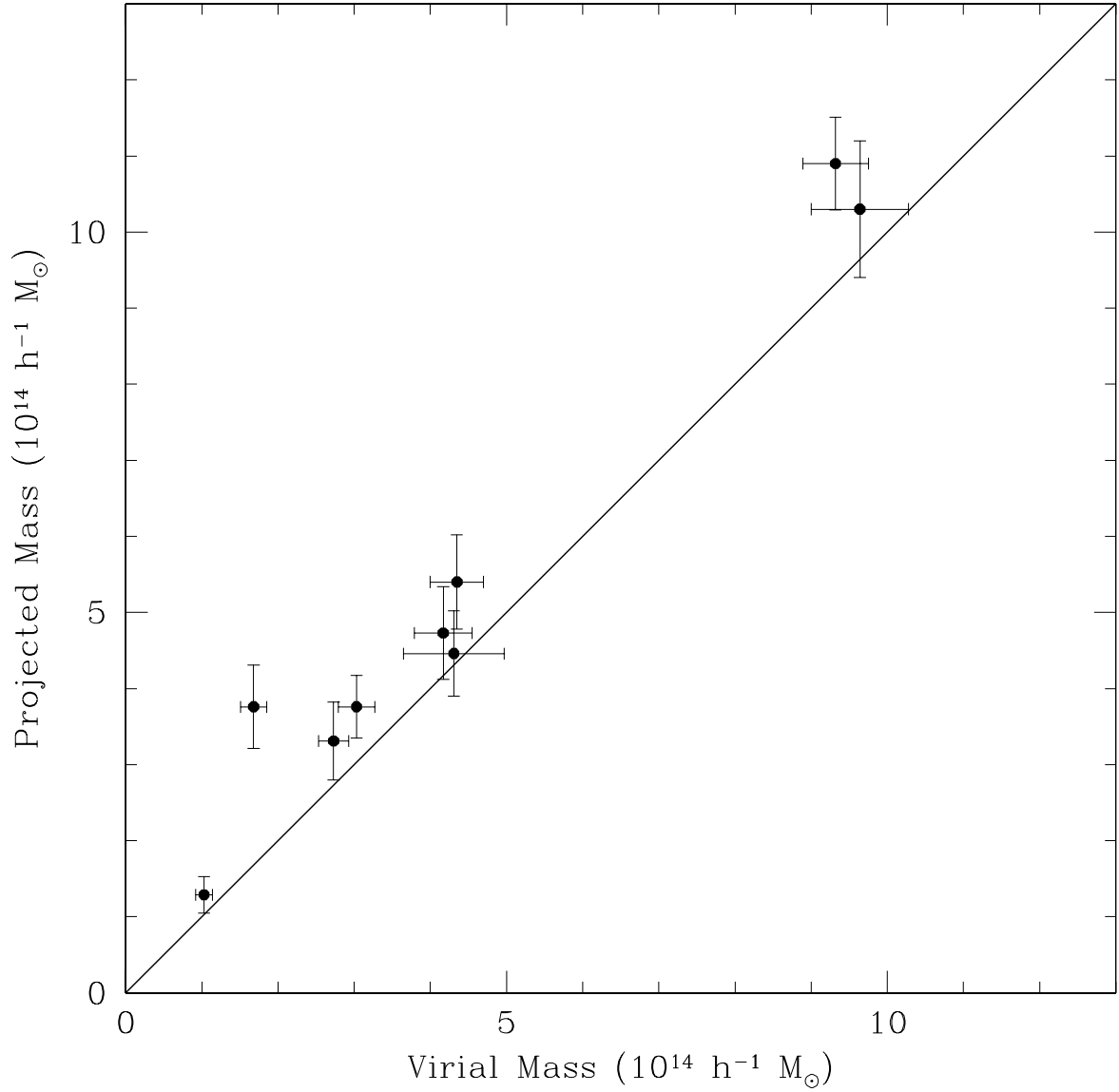} 
\caption{\label{vp} Comparison of projected mass estimates to virial
mass estimates.  The solid line has a slope of
unity and intercepts the origin.} 
\end{figure}

\begin{figure}
\figurenum{10}
\plotone{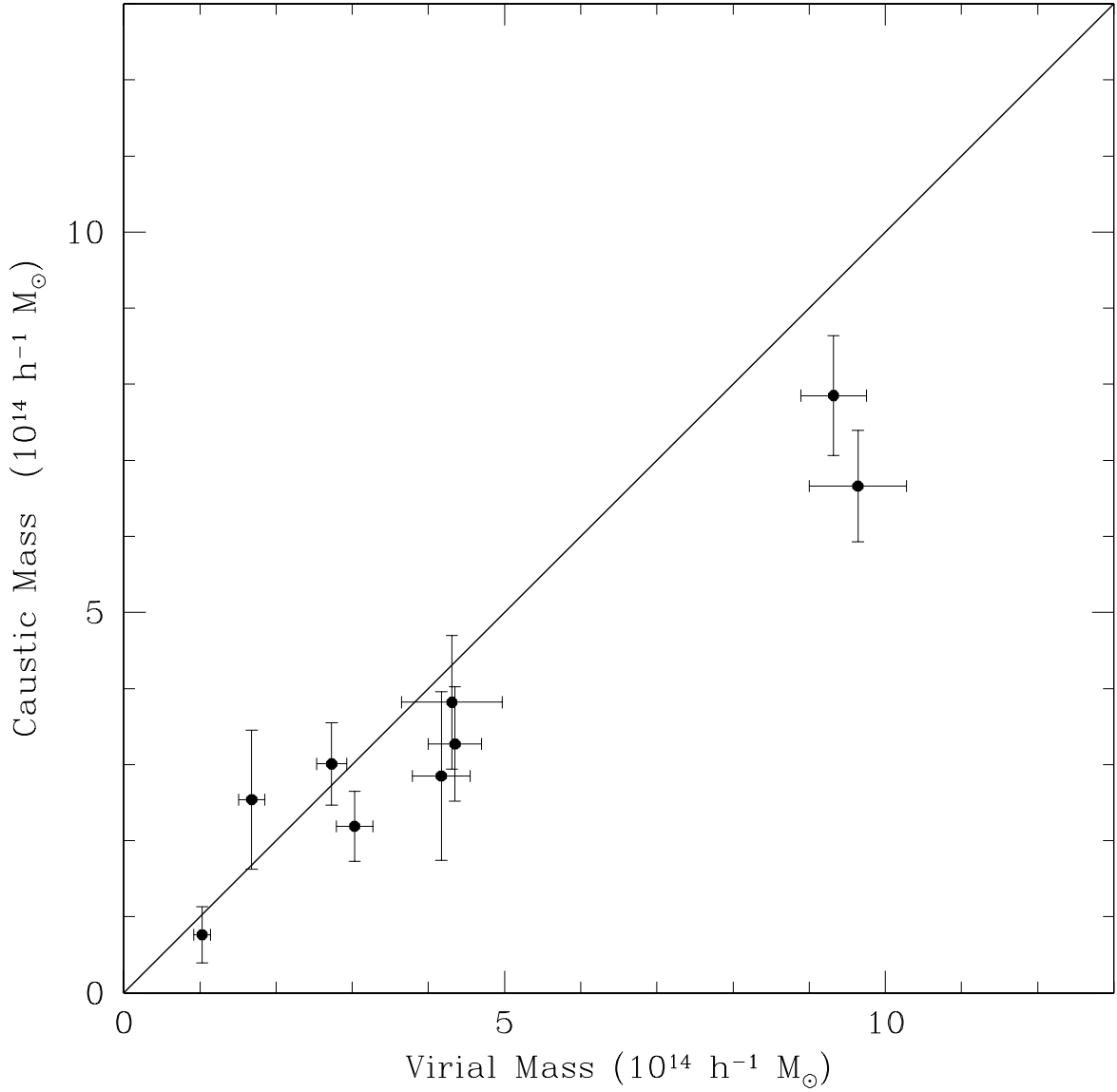} 
\caption{\label{vc} Comparison of caustic mass estimates to estimates
based on the virial theorem.  The solid line has a slope of
unity and intercepts the origin.} 
\end{figure}

\begin{figure}
\figurenum{11}
\plotone{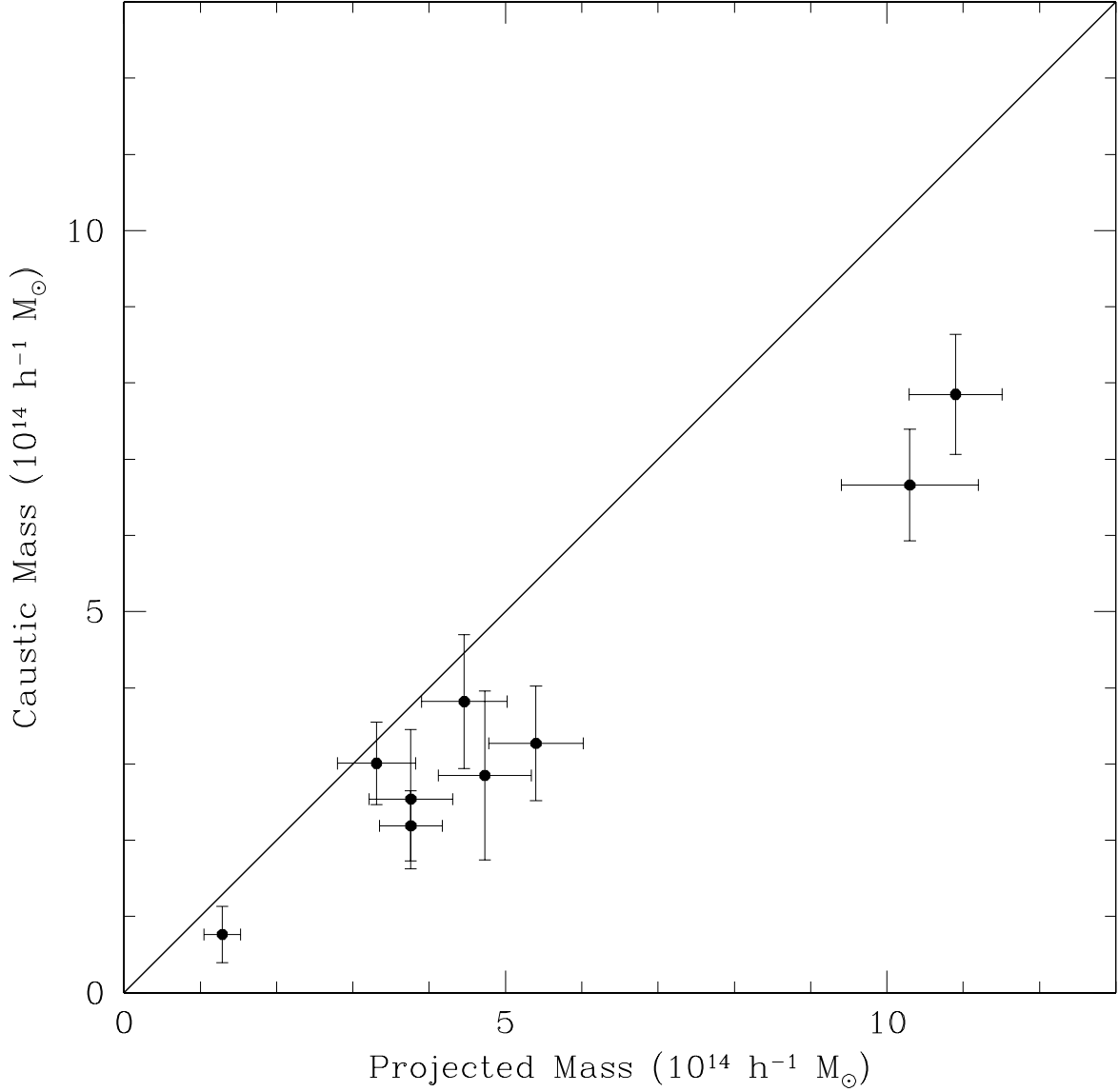} 
\caption{\label{pc} Comparison of caustic mass estimates to estimates
based on the projected mass estimator.  The solid line has a slope of
unity and intercepts the origin.} 
\end{figure}

\begin{figure}
\figurenum{12}
\plotone{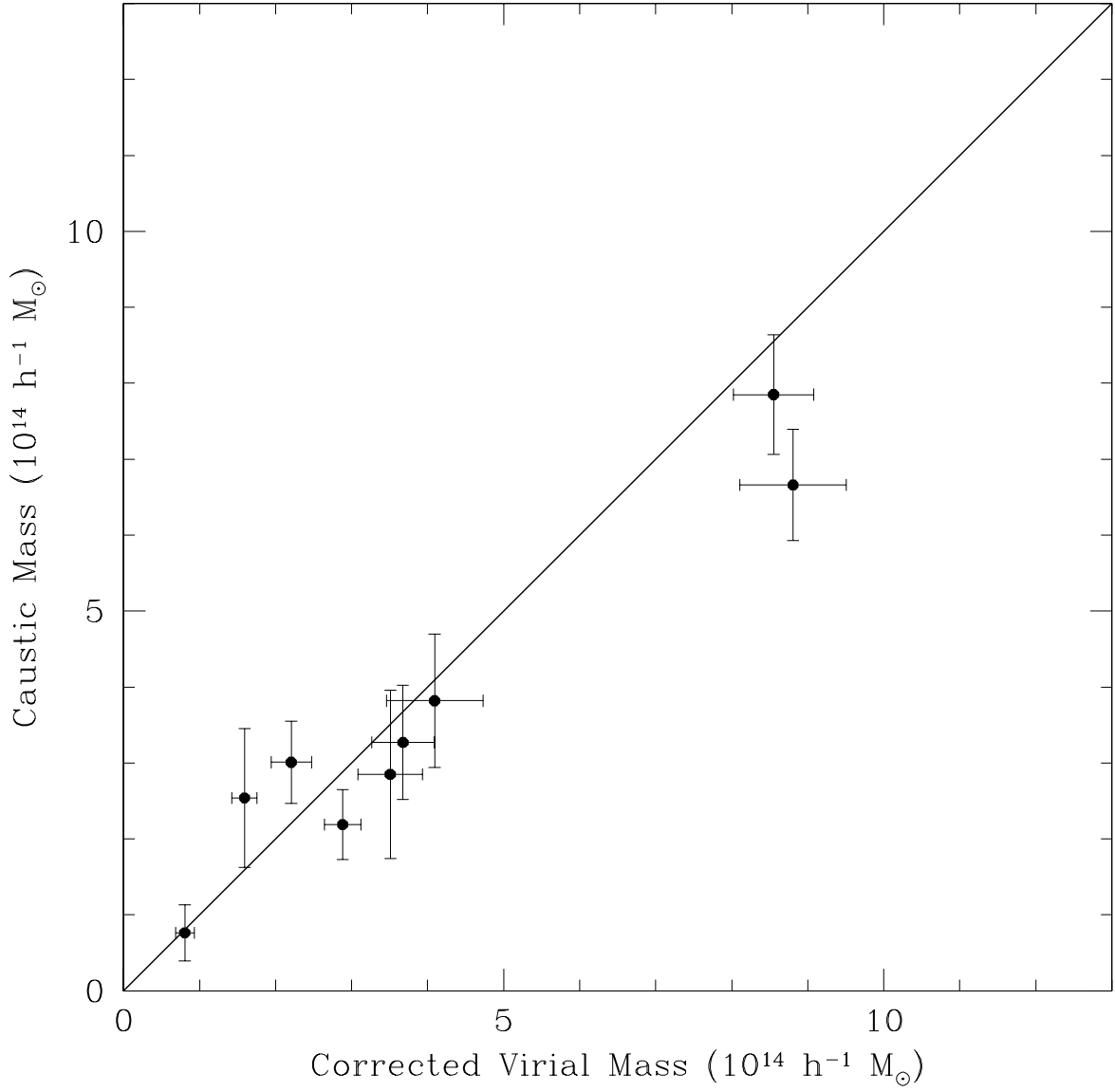} 
\caption{\label{cvc} Comparison of caustic mass estimates to estimates
based on the virial mass estimator with a correction for the surface
pressure term.  The solid line has a slope of
unity and intercepts the origin.} 
\end{figure}

\clearpage

\begin{figure}
\figurenum{13}
\plotone{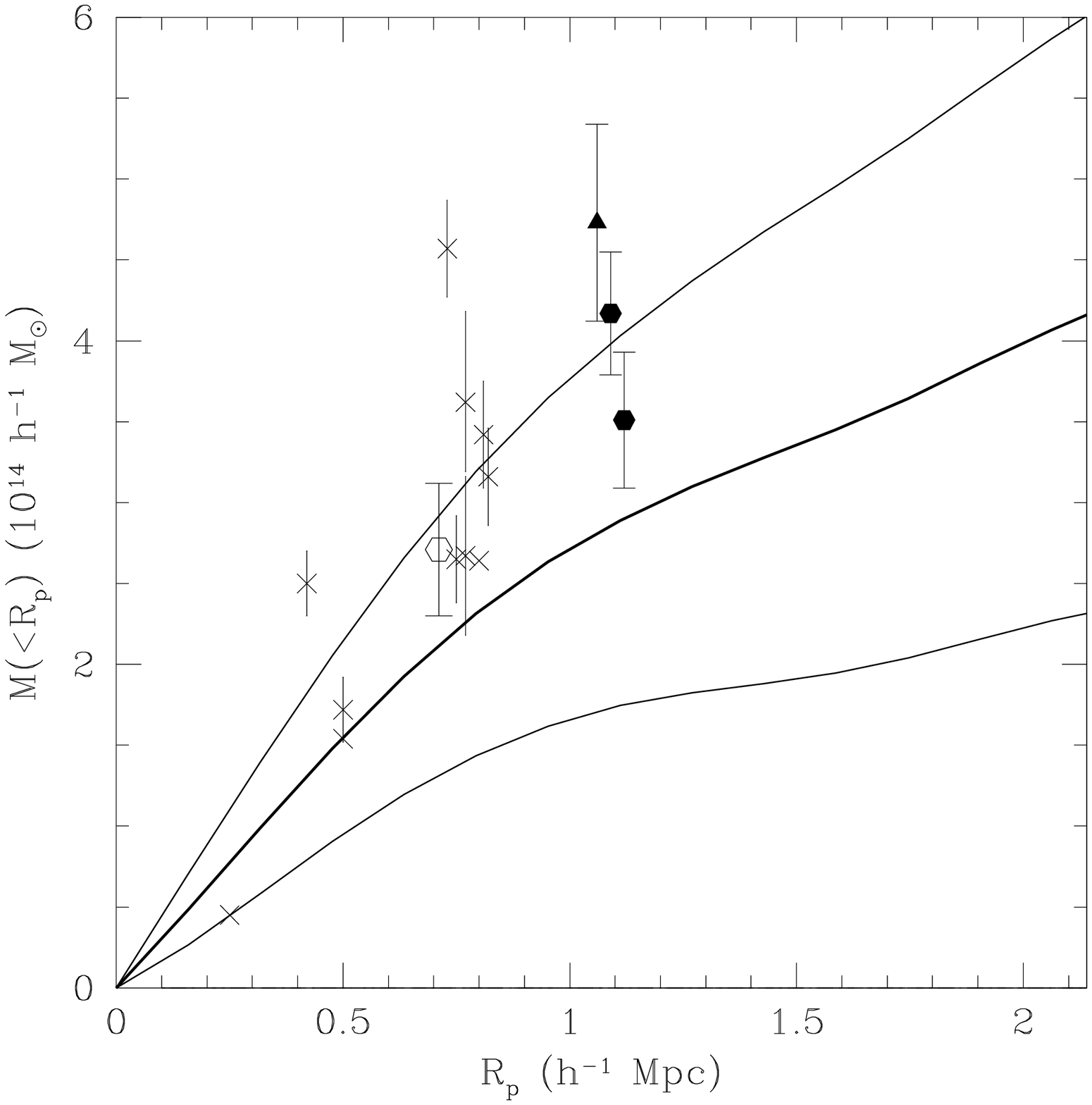} 
\caption{\label{a119mp} Mass profile of A119. Thick and thin solid
lines show the caustic mass profile and the 1-$\sigma$ uncertainties.
Crosses show X-ray estimates, the open hexagon is the estimate from
the $M_{500}-T_X$ relation, the triangle is the projected mass, and
the filled hexagons show the virial mass with and without correction
for the surface pressure term.}
\end{figure}

\begin{figure}
\figurenum{14}
\plotone{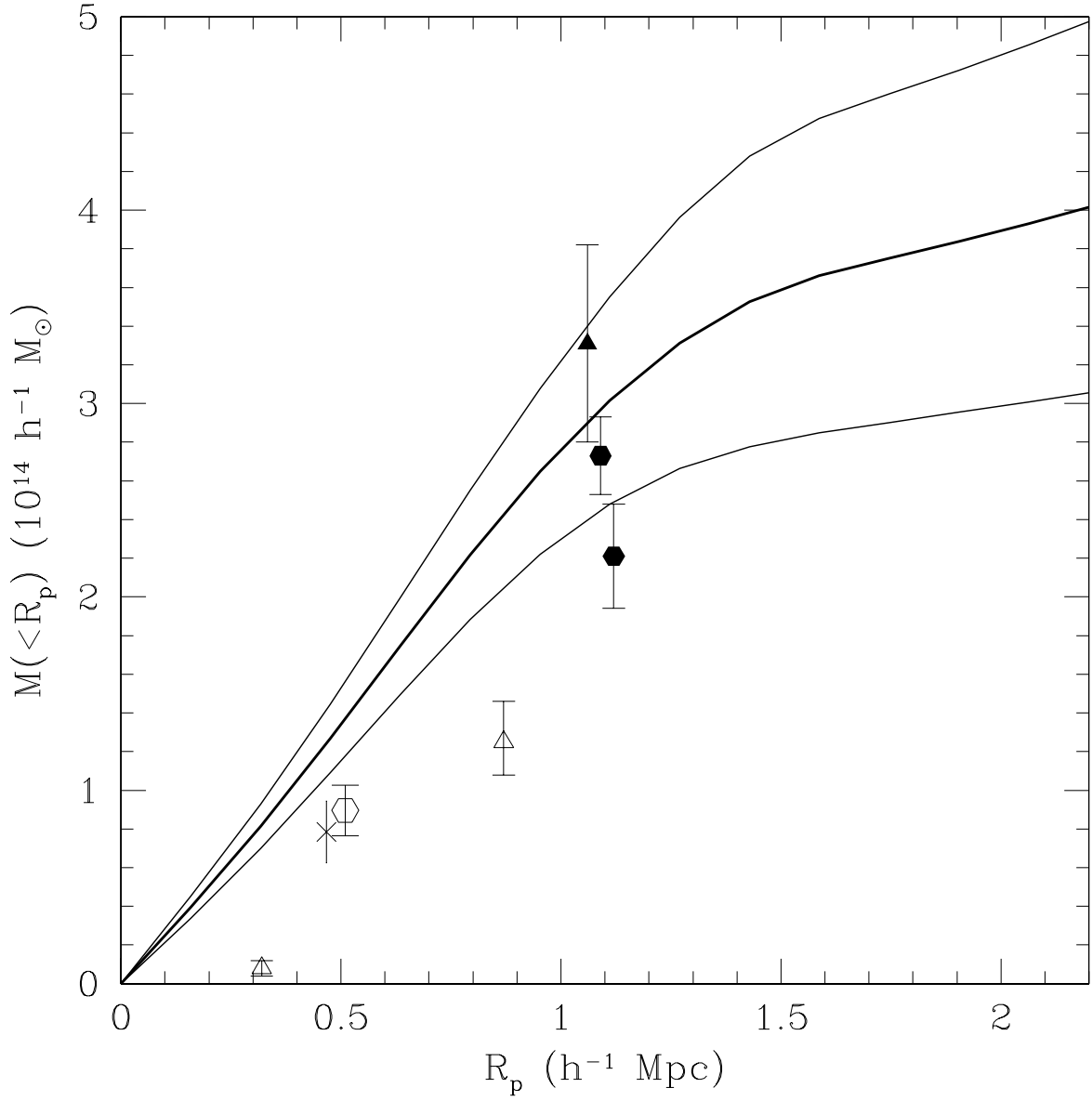} 
\caption{\label{a168mp} Same as Figure \ref{a119mp} for A168.  The
open triangles indicate the mass estimates of \citet{girardi98}.  See
text for details.} 
\end{figure}

\begin{figure}
\figurenum{15}
\plotone{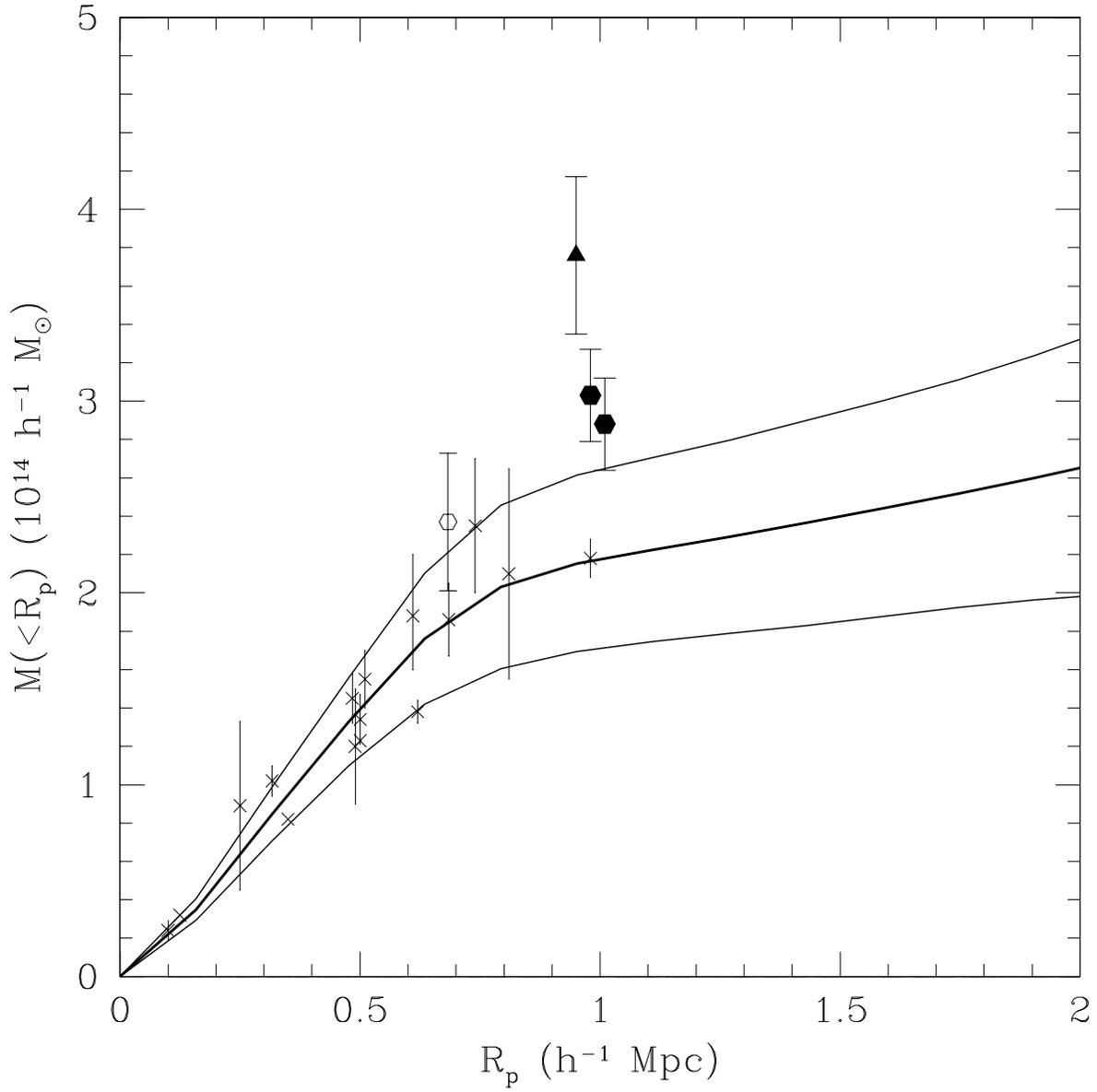} 
\caption{\label{a496mp} Same as Figure \ref{a119mp} for A496.} 
\end{figure}

\begin{figure}
\figurenum{16}
\plotone{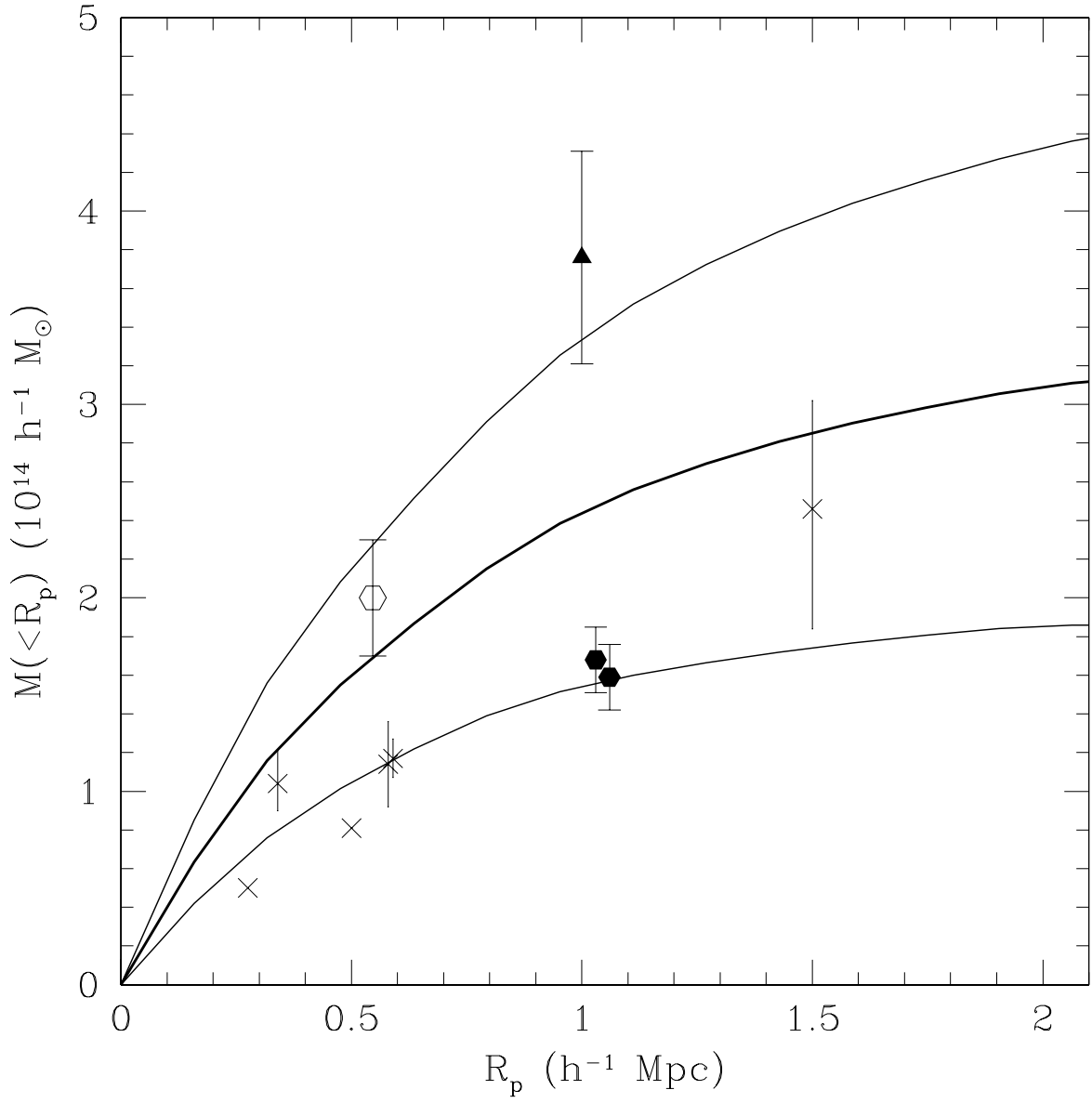} 
\caption{\label{a539mp} Same as Figure \ref{a119mp} for A539.} 
\end{figure}

\begin{figure}
\figurenum{17}
\plotone{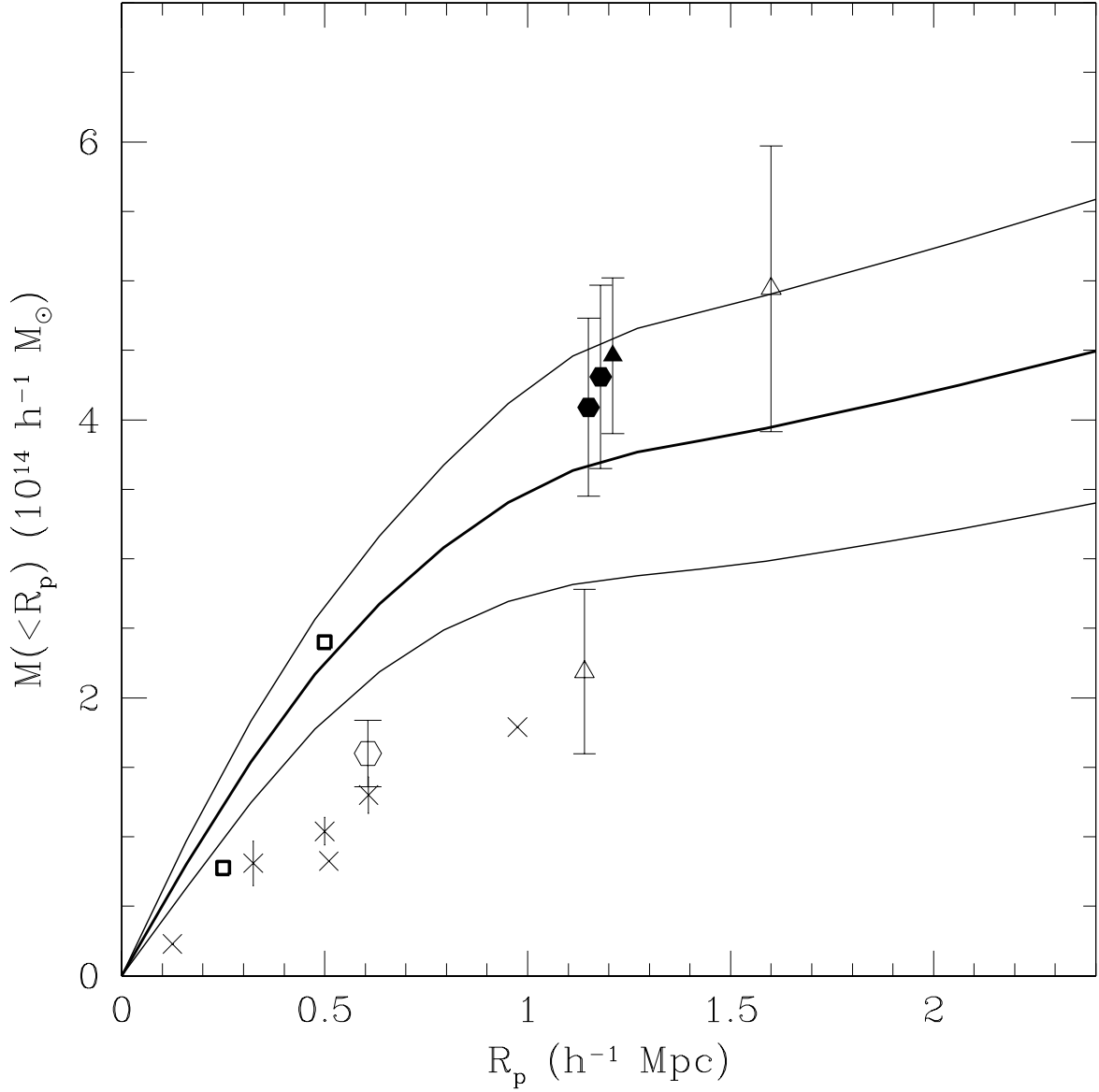} 
\caption{\label{a1367mp} Same as Figure \ref{a119mp} for A1367. The
open squares show the sum of the two mass components studied in
\citet{hanka1367}. The open triangles show the mass estimates of
\citet{girardi98}.  See text for details.} 
\end{figure}

\begin{figure}
\figurenum{18}
\plotone{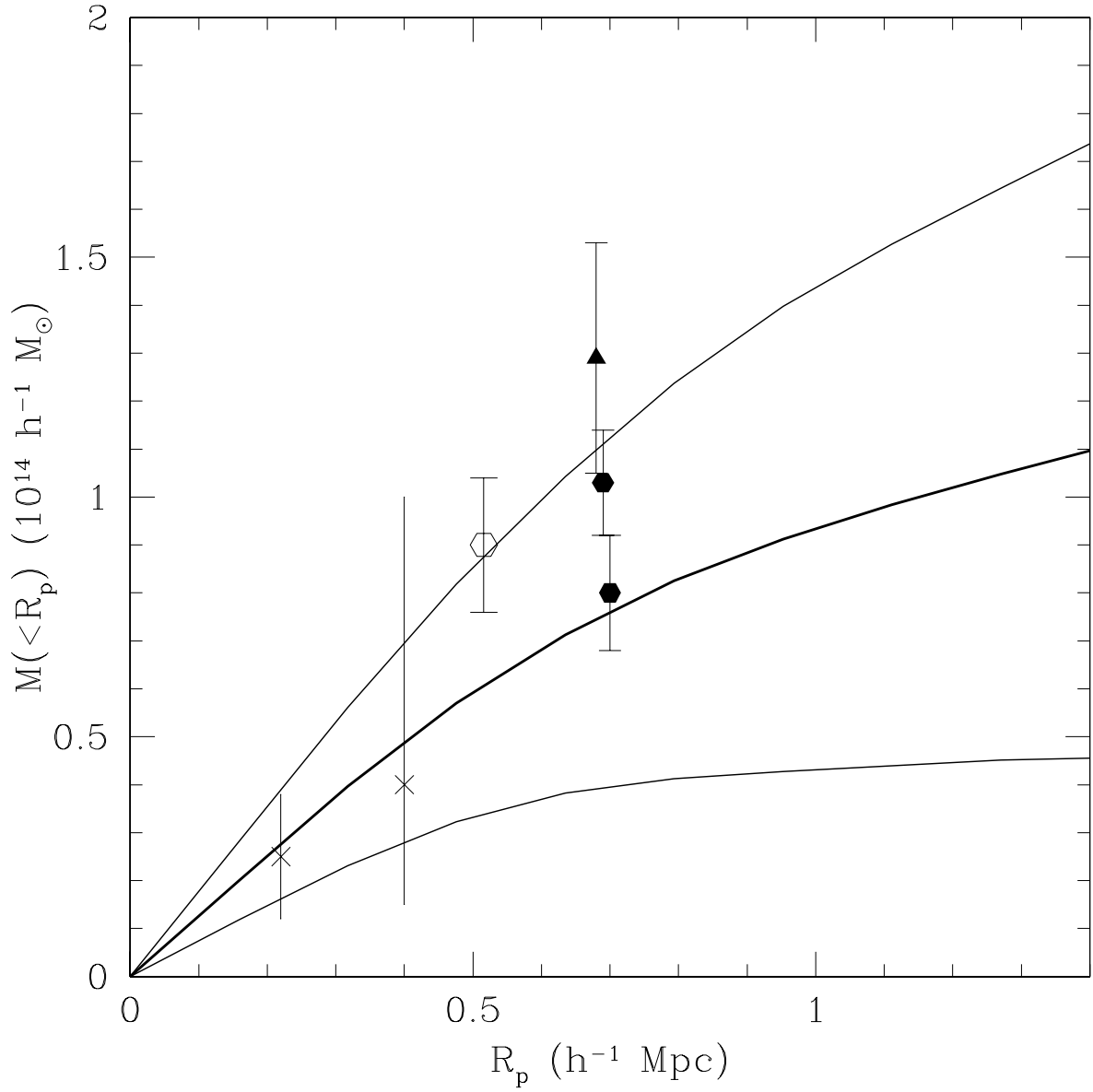} 
\caption{\label{a194mp} Same as Figure \ref{a119mp} for A194.} 
\end{figure}

\begin{table*}[th] \footnotesize
\begin{center}
\caption{\label{sample} \sc CAIRNS Basic Properties}
\begin{tabular}{lcccrrrrcc}
\tableline
\tableline
\tablewidth{0pt}
Cluster &\multicolumn{2}{c}{X-ray Coordinates} & $cz_\odot$ & $\sigma
 _p (3\sigma)$ & $\sigma_p (caustics)$ & $L_X /10^{43}$ & $T_X$ & Richness  \\ 
 & RA (J2000) & DEC (J2000) &  $\kms$ & $\kms$ & $\kms$ & erg~s$^{-1}$& keV  \\ 
\tableline
A119 & 00 56 12.9 & -01 14 06 & 13268 &698$^{+36}_{-31}$ & 619$^{+44}_{-36}$ & 8.1 & 5.1 & 1  \\
A168 & 01 15 08.8 & +00 21 14 & 13395 &579$^{+36}_{-30}$ & 531$^{+39}_{-32}$ & 2.7  & 2.6 & 2  \\
A496 & 04 33 35.2 & -13 14 45 & 9900 &721$^{+35}_{-30}$ & 598$^{+36}_{-30}$ & 8.9  & 4.7 & 1  \\
A539 & 05 16 32.1 & +06 26 31 & 8717 &734$^{+53}_{-44}$ & 717$^{+60}_{-48}$ & 2.7  & 3.0 & 1 \\ 
A576 & 07 21 31.6 & +55 45 50 & 11510 &1009$^{+41}_{-36}$ & 881$^{+44}_{-38}$ & 3.5 & 3.7 & 1  \\
A1367 & 11 44 36.2 & +19 46 19 & 6495 &782$^{+56}_{-46}$ & 745$^{+52}_{-43}$ & 4.1  & 3.5 & 2  \\ 
A1656 & 12 59 31.9 & +27 54 10 & 6973 &1042$^{+33}_{-30}$ & 957$^{+30}_{-28}$ & 18.0  & 8.0 & 2  \\
A2199 & 16 28 39.5 & +39 33 00 & 9101 &796$^{+38}_{-33}$ & 722$^{+37}_{-32}$ & 9.1  & 4.7 & 2  \\
\tableline
A194 & 01 25 50.4 & -01 21 54 & 5341 &495$^{+41}_{-33}$ & 402$^{+38}_{-29}$ & 0.4  & 2.6 & 0  \\
\tableline
\end{tabular}
\end{center}
\end{table*}

\begin{table*}[th] \footnotesize
\begin{center}
\caption{\label{catalogs} \sc CAIRNS Redshift Catalogs}
\begin{tabular}{lrrrrrr}
\tableline
\tableline
\tablewidth{0pt}
Cluster & $R_{max}$ & $cz_\odot$ & $\sigma _p$ & $N_{cz}$ & $N_{CAIRNS}$ &
 $N_{mem}$ \\ 
 & Degrees & $\kms$ & $\kms$ &  &  & \\ 
\tableline
A119 & 5 & 13268$\pm$47 &619$^{+44}_{-36}$ & 3381\tablenotemark{a} & 669\tablenotemark{a} & 121 \\
A168 & 5 & 13395$\pm$46 &531$^{+39}_{-32}$ & 3557\tablenotemark{a}  & 525\tablenotemark{a} & 117 \\
A496 & 5 & 9900$\pm$46 &598$^{+36}_{-30}$ & 1043  & 515 & 169 \\
A539 & 10 & 8717$\pm$68 &717$^{+60}_{-48}$ & 796 & 360 & 91 \\ 
A576 & 5 & 11510$\pm$54 &881$^{+44}_{-38}$ & 1209 & 529 & 233 \\
A1367 & 10 & 6495$\pm$71 &745$^{+52}_{-43}$ & 1864 & 502 & 128 \\ 
A1656 & 10 & 6973$\pm$45 &957$^{+30}_{-28}$ & 4159 & 1239  & 548 \\
A2199 & 6.5 & 9101$\pm$50 &722$^{+37}_{-32}$ & 1539 & 820 & 218 \\
\tableline
A194 & 10 & 5341$\pm$51 &402$^{+38}_{-29}$ & 6916\tablenotemark{a}  & -- & 75  \\
\tableline
\tablenotetext{a}{Overlapping samples. See text for details.}
\end{tabular}
\end{center}
\end{table*}

\begin{deluxetable}{lccccc}  
\tablecolumns{6}  
\tablewidth{0pc}  
\tablecaption{Spectroscopic Data for A119 and A168\tablenotemark{a}\label{a119a168cz}}  
\small
\tablehead{  
\colhead{}    RA & DEC & $cz$ & $\sigma _{cz}$ & Reference & Cluster \\
\colhead{}    (J2000) & (J2000) & ($\kms$)  & ($\kms$) & & \\
}
\startdata
 00:36:29.23 & -00:43:43.0 & 10214 & 051 & 3 & A119 \\ 
 00:36:34.18 & -01:04:46.6 & 33134 & 036 & 3 & A119 \\ 
 00:36:45.36 & -00:40:00.5 & 32774 & 040 & 3 & A119 \\ 
 00:36:48.36 & -00:37:27.1 & 48774 & 046 & 3 & A119 \\ 
 00:36:54.22 & -00:06:39.2 & 51401 & 019 & 3 & A119 \\ 
\enddata	     						  
\tablenotetext{a}{The complete version of this table is in the
 electronic edition of the Journal.  The printed edition contains only
 a sample.} 
\tablerefs{
(1) FAST spectra; (2) NED; (3) SDSS.}
\end{deluxetable}

\begin{deluxetable}{lccccc}  
\tablecolumns{6}  
\tablewidth{0pc}  
\tablecaption{Spectroscopic Data for A194\tablenotemark{a}\label{a194cz}}  
\small
\tablehead{  
\colhead{}    RA & DEC & $cz$ & $\sigma _{cz}$ & Reference & Cluster \\
\colhead{}    (J2000) & (J2000) & ($\kms$)  & ($\kms$) & & \\
}
\startdata
 00:45:40.45 & -00:52:16.6 & 32739 & 050 & 2 & A119 \\ 
 00:45:44.11 & -00:50:37.6 & 31930 & 037 & 2 & A119 \\ 
 00:45:44.78 & -01:05:20.8 & 20175 & 012 & 2 & A119 \\ 
 00:45:46.72 & -01:06:40.3 & 35461 & 042 & 2 & A119 \\ 
 00:45:47.95 & -00:51:10.1 & 32949 & 040 & 2 & A119 \\ 
\enddata	     						  
\tablenotetext{a}{The complete version of this table is in the
 electronic edition of the Journal.  The printed edition contains only
 a sample.} 
\tablerefs{
(1) FAST spectra; (2) NED.}
\end{deluxetable}

\begin{deluxetable}{lcccc}  
\tablecolumns{5}  
\tablewidth{0pc}  
\tablecaption{Spectroscopic Data for A496\tablenotemark{a}\label{a496cz}}  
\small
\tablehead{  
\colhead{}    RA & DEC & $cz$ & $\sigma _{cz}$  & Reference\\
\colhead{}    (J2000) & (J2000) & ($\kms$)  & ($\kms$) & \\
}
\startdata
 04:14:01.49 & -13:23:23.8 & 09337 & 000 & 02 \\ 
 04:14:08.57 & -13:08:34.9 & 09622 & 032 & 01 \\ 
 04:14:32.78 & -13:11:02.4 & 08656 & 040 & 01 \\ 
 04:14:36.00 & -13:10:29.7 & 09098 & 038 & 02 \\ 
 04:14:36.08 & -13:19:01.6 & 08566 & 016 & 01 \\ 
\enddata	     						  
\tablenotetext{a}{The complete version of this table is in the
 electronic edition of the Journal.  The printed edition contains only
 a sample.} 
\tablerefs{
(1) FAST spectra; (2) NED.}
\end{deluxetable}

\begin{deluxetable}{lcccc}  
\tablecolumns{5}  
\tablewidth{0pc}  
\tablecaption{Spectroscopic Data for A539\tablenotemark{a}\label{a539cz}}  
\small
\tablehead{  
\colhead{}    RA & DEC & $cz$ & $\sigma _{cz}$  & Reference\\
\colhead{}    (J2000) & (J2000) & ($\kms$)  & ($\kms$) & \\
}
\startdata
 04:38:58.98 &  05:37:11.3 & 08310 & 005 & 2  \\ 
 04:39:00.60 &  07:16:05.0 & 73509 & 150 & 2  \\ 
 04:39:02.26 &  05:20:43.7 & 62357 & 300 & 2  \\ 
 04:39:45.10 &  03:01:57.0 & 04518 & 030 & 2  \\ 
 04:39:51.50 &  07:03:19.0 & 04693 & 007 & 2  \\ 
\enddata	     						  
\tablenotetext{a}{The complete version of this table is in the
 electronic edition of the Journal.  The printed edition contains only
 a sample.} 
\tablerefs{
(1) FAST spectra; (2) NED.}
\end{deluxetable}

\begin{deluxetable}{lcccc}  
\tablecolumns{5}  
\tablewidth{0pc}  
\tablecaption{Spectroscopic Data for A1367\tablenotemark{a}\label{a1367cz}}  
\small
\tablehead{  
\colhead{}    RA & DEC & $cz$ & $\sigma _{cz}$  & Reference\\
\colhead{}    (J2000) & (J2000) & ($\kms$)  & ($\kms$) & \\
}
\startdata
 11:03:25.39 &  18:08:12.2 & 00979 & 004 & 2  \\ 
 11:03:38.70 &  19:19:43.1 & 09174 & 050 & 2  \\ 
 11:04:17.20 &  18:03:54.0 & 56961 & 000 & 2  \\ 
 11:04:32.20 &  17:07:40.1 & 13028 & 061 & 2  \\ 
 11:04:36.20 &  21:24:18.0 & 56241 & 087 & 2  \\ 
\enddata	     						  
\tablenotetext{a}{The complete version of this table is in the
 electronic edition of the Journal.  The printed edition contains only
 a sample.} 
\tablerefs{
(1) FAST spectra; (2) NED.}
\end{deluxetable}

\begin{deluxetable}{lcccc}  
\tablecolumns{4}  
\tablewidth{0pc}  
\tablecaption{Spectroscopic Data for Coma\tablenotemark{a}\label{comacz}}  
\small
\tablehead{  
\colhead{}    RA & DEC & $cz$ & $\sigma _{cz}$ & Reference \\
\colhead{}    (J2000) & (J2000) & ($\kms$)  & ($\kms$) &\\
}
\startdata
 12:15:03.51 &  29:06:02.3 & 31301 & 38 & 2  \\ 
 12:15:06.90 &  29:01:10.0 & 7445 & 72 & 2  \\ 
 12:15:15.77 &  28:50:31.0 & 57063 & 67 & 2  \\ 
 12:15:16.12 &  29:15:07.1 & 40917 & 66 & 2  \\ 
 12:15:23.17 &  26:53:05.6 & 7518 & 24 & 1  \\ 
\enddata	     						  
\tablenotetext{a}{The complete version of this table is in the
 electronic edition of the Journal.  The printed edition contains only
 a sample.} 
\end{deluxetable}

\begin{table*}[th] \footnotesize
\begin{center}
\caption{\label{centers} \sc CAIRNS Hierarchical Centers}
\begin{tabular}{lcccr}
\tableline
\tableline
\tablewidth{0pt}
Cluster &\multicolumn{2}{c}{Hierarchical Center} & $cz_{cen}$ & $\Delta R$  \\ 
 & RA (J2000) & DEC (J2000) & $\kms$ & $\kpc$  \\ 
\tableline
A119 & 00 56 10.1 & -01 15 20 & 13278 & 56  \\
A168 & 01 15 00.7 & +00 15 31 & 13493 & 239  \\
A496 & 04 33 38.6 & -13 15 47 & 9831 & 24  \\
A539 & 05 16 37.0 & +06 26 57 & 8648 & 33 \\ 
A576 & 07 21 32.0 & +55 45 21 & 11487 & 16  \\
A1367 & 11 44 49.1 & +19 46 03 & 6509 & 61  \\ 
A1656 & 13 00 00.7 & +27 56 51 & 7093 & 153  \\
A2199 & 16 28 47.0 & +39 30 22 & 9156 & 86  \\
\tableline
A194 & 01 25 48.0 & -01 21 34 & 5317 & 11  \\
\tableline
\end{tabular}
\end{center}
\end{table*}

\begin{table*}[th] \footnotesize
\begin{center}
\caption{\label{radii} \sc CAIRNS Virial and Turnaround Radii}
\begin{tabular}{lccrrr}
\tableline
\tableline
\tablewidth{0pt}
Cluster & $r_{200}$ & $r_t$ & $M_{200}$ & $M_{t}$ & $M_{t}/M_{200}$ \\ 
 & $\Mpc$ & $\Mpc$ & $10^{13} M_\odot$ & $10^{13} M_\odot$ &  \\ 
\tableline
A119  & 1.07 & 5.4 & 28.5 & 63 & 2.2 \\
A168  & 1.09 & 5.5 & 30.1 & 66 & 2.2 \\
A496  & 0.98 & 4.2 & 21.9 & 30 & 1.4 \\
A539  & 1.03 & 4.3 & 25.4 & 32 & 1.2 \\ 
A576  & 1.42 & 6.0 & 66.6 & 88 & 1.3 \\
A1367 & 1.18 & 5.2 & 38.2 & 56 & 1.5 \\ 
A1656 & 1.50 & 7.4 & 78.5 & 165 & 2.1 \\
A2199 & 1.12 & 5.3 & 32.7 & 58 & 1.8 \\
\tableline
A194  & 0.69 & 3.3 &  7.6 & 15 & 2.0 \\
\tableline
\end{tabular}
\end{center}
\end{table*}

\begin{table*}[th] \footnotesize
\begin{center}
\caption{\label{mpfitsci} \sc CAIRNS Mass Profile Fit Parameters}
\begin{tabular}{lccccrccc}
\tableline
\tableline
\tablewidth{0pt}
Cluster & Form & $R_{max}$ & $a$ & $1 \sigma$ & $M(a)$ & $1\sigma$ & $\chi ^2$ & $\nu$ \\ 
 & & $\Mpc$ & $\Mpc$ & $\Mpc$ & $10^{13} M_\odot$ & $10^{13} M_\odot$ & &  \\ 
\tableline
A119 & NFW & 5.4 & 0.17 & 0.10-0.27 & 4.9 & 3.7-6.4 & 0.04 & 32 \\
     & Hernquist & 5.4 & 0.58 & 0.44-0.62 & 17.5 & 15.0-21.5 & 0.51 & 32 \\
     & SIS & 5.4 & 0.50 & -- & 8.4 & 7.8-9.2 & 12.5 & 33 \\
A168 & NFW & 5.5 & 0.21 & 0.17-0.25 & 5.2 & 4.6-5.8 & 2.39 & 32 \\
     & Hernquist & 5.5 & 0.65 & 0.57-0.71 & 18.0 & 16.4-19.2 & 1.97 & 32 \\
     & SIS & 5.5 & 0.50 & -- & 8.2 & 8.0-8.6 & 67.8 & 33 \\
A496 & NFW & 4.0 & 0.07 & 0.05-0.10 & 2.1 & 1.8-2.5 & 2.67 & 24 \\
     & Hernquist & 4.0 & 0.31 & 0.26-0.38 & 8.9 & 8.2-9.9 & 0.73 & 24 \\
     & SIS & 4.0 & 0.50 & -- & 6.2 & 5.9-6.6 & 62.6 & 25 \\
A539 & NFW & 2.5 & 0.07 & 0.04-0.13 & 2.4 & 1.8-3.5 & 0.13 & 15 \\
     & Hernquist & 2.5 & 0.25 & 0.17-0.44 & 9.5 & 7.9-12.6 & 0.16 & 15 \\
     & SIS & 2.5 & 0.50 & -- & 9.3 & 7.4-10.8 & 12.4 & 16 \\
A576 & NFW & 4.3 & 0.13 & 0.12-0.14 & 7.4 & 6.9-7.9 & 7.60 & 26 \\
     & Hernquist & 4.3 & 0.43 & 0.40-0.46 & 26.9 & 25.7-28.0 & 1.64 & 26 \\
     & SIS & 4.3 & 0.50 & -- & 15.7 & 15.4-16.1 & 189 & 27 \\
A1367 & NFW & 5.2 & 0.07 & 0.05-0.09 & 3.3 & 2.9-3.7 & 1.37 & 31 \\
     & Hernquist & 5.2 & 0.29 & 0.25-0.34 & 14.5 & 13.5-15.5 & 0.88 & 31 \\
     & SIS & 5.2 & 0.50 & -- & 8.1 & 7.7-8.4 & 99.3 & 32 \\
A1656 & NFW & 7.4 & 0.15 & 0.14-0.17 & 10.3 & 9.7-11.0 & 2.46 & 45 \\
     & Hernquist & 7.4 & 0.53 & 0.48-0.55 & 38.6 & 36.7-40.2 & 22.4 & 45 \\
     & SIS & 7.4 & 0.50 & -- & 18.6 & 18.3-19.1 & 292 & 46 \\
A2199 & NFW & 4.1 & 0.15 & 0.11-0.19 & 4.7 & 4.0-5.5 & 0.50 & 25 \\
     & Hernquist & 4.1 & 0.47 & 0.39-0.57 & 16.7 & 15.0-18.5 & 1.21 & 25 \\
     & SIS & 4.1 & 0.50 & -- & 9.8 & 9.3-10.2 & 30.8 & 26 \\
\tableline
A194 & NFW & 3.3 & 0.11 & 0.06-0.19 & 1.3 & 0.9-1.8 & 0.01 & 19 \\
     & Hernquist & 3.3 & 0.35 & 0.24-0.52 & 4.5 & 3.4-5.8 & 0.26 & 19 \\
     & SIS & 3.3 & 0.50 & -- & 3.7 & 3.2-4.1 & 6.01 & 20 \\
\tableline
\end{tabular}
\end{center}
\end{table*}

\begin{table*}[th] \footnotesize
\begin{center}
\caption{\label{virial} \sc CAIRNS Virial and Projected Masses}
\begin{tabular}{lcrrrr}
\tableline
\tableline
\tablewidth{0pt}
Cluster & $r_{200}$ & $M_{200}$ & $M_{proj}$ & $M_{vir}$ & $M_{cv}$ \\ 
 & $\Mpc$  & $10^{13} M_\odot$ & $10^{13} M_\odot$   & $10^{13} M_\odot$ & $10^{13} M_\odot$  \\ 
\tableline
A119  & 1.07& 28.5 & 47.3$\pm$6.1 &  41.7$\pm$3.8 & 35.1$\pm$4.2  \\
A168  & 1.09& 30.1 & 33.1$\pm$5.1 &  27.3$\pm$2.0 & 22.1$\pm$2.7    \\
A496  & 0.98& 21.9 & 37.6$\pm$4.1 &  30.3$\pm$2.4 & 28.8$\pm$2.4  \\
A539  & 1.03& 25.4 & 37.6$\pm$5.5 &  16.8$\pm$1.7 & 15.9$\pm$1.7  \\ 
A576  & 1.42& 66.6 & 103.0$\pm$9.0 &  96.4$\pm$6.4 & 88.0$\pm$7.0    \\
A1367 & 1.18& 38.2 & 44.6$\pm$5.6 &  43.1$\pm$6.6 & 40.9$\pm$6.4  \\ 
A1656 & 1.50& 78.5 & 109.0$\pm$6.1 &  93.2$\pm$4.3 & 85.5$\pm$5.3   \\
A2199 & 1.12& 32.7 & 54.0$\pm$6.2 &  43.5$\pm$3.5 & 36.8$\pm$4.1   \\
\tableline
A194  & 0.69&  7.6 & 12.9$\pm$2.4 &  10.3$\pm$1.1 & 8.0$\pm$1.2  \\
\tableline
\end{tabular}
\end{center}
\end{table*}

\end{document}